\documentclass[twocolumn, prx, aps, superscriptaddress, floatfix, nofootinbib,10pt]{revtex4-2}

\usepackage[utf8]{inputenc}
\usepackage[english]{babel}
\usepackage[sort&compress]{natbib}

\usepackage{graphicx}
\usepackage{amsmath, amssymb, amsthm}
\usepackage{tensor}
\usepackage{braket}
\usepackage{cases}
\usepackage{dsfont}
\usepackage[dvipsnames]{xcolor}
\usepackage{hyperref}
\usepackage{nameref}
\hypersetup{
	colorlinks = true,
	allcolors = RoyalPurple
}
\usepackage[percent]{overpic}
\usepackage{svg}

\usepackage{tikz}
\usepackage{circuitikz}
\usetikzlibrary{quantikz}
\usetikzlibrary{shapes}

\usepackage[export]{adjustbox}
\usepackage{leftidx}
\usepackage{BOONDOX-cal} 
\usepackage{orcidlink}

\DeclareMathOperator{\sgn}{sgn}
\DeclareMathOperator{\Tr}{Tr}

\DeclareMathOperator{\Span}{span}
\DeclareMathOperator{\rank}{rank}

\newtheorem{definition}{Definition}[]

\newtheorem{lemma}[definition]{Lemma}
\newtheorem{theorem}[definition]{Proposition}

\renewcommand{\braket}[1]{\langle #1\rangle}
\renewcommand{\ket}[1]{\vert #1 \rangle}
\renewcommand{\bra}[1]{\langle #1 \vert}

\newcommand{\norm}[1]{\left\lVert #1 \right\rVert}
\newcommand{\abs}[1]{\left\lvert #1 \right\rvert}

\renewcommand{\epsilon}{\varepsilon}

\definecolor{BLUE}{RGB}{0,0,255}

\newcommand{\change}[1]{#1}
\newcommand{\changeTwo}[1]{#1}

\newcommand{\tomo}[3]{
	\begin{tikzpicture}
		\node[rectangle,minimum width=0.5cm,minimum height=1cm,draw] at (0,0) {$U$};
		\draw [thick] (-0.5,0.25) -- (-0.25,0.25)
		(-0.5,-0.25) -- (-0.25,-0.25)
		(-0.5,0.25) -- (-0.5,0.75)
		(-0.5,-0.25) -- (-0.5,-0.75)
		(-0.5,0.75) -- (0.75,0.75)
		(-0.5,-0.75) -- (0.75,-0.75)
		(0.25,0.25) -- (0.75,0.25)
		(0.25,-0.25) -- (0.75,-0.25)
		(0.5,-0.2) -- (0.6,-0.3)
		(0.5,-0.7) -- (0.6,-0.8);
		\node at (-1,0.5) {#1};
		\node at (-1,-0.5) {#2};
		\node at (1,0.5) {#3};
	\end{tikzpicture}
}

\newcommand{\tomoblock}[5]{
	\begin{tikzpicture}
		\node[rectangle,minimum width=0.5cm,minimum height=1.5cm,draw] at (0,0) {$U$};
		\draw [thick] (-0.5,0.5) -- (-0.25,0.5)
		(-0.5,-0.5) -- (-0.25,-0.5)
		(-0.5,0.5) -- (-0.5,1)
		(-0.5,-0.5) -- (-0.5,-1)
		(-0.5,1) -- (0.75,1)
		(-0.5,-1) -- (0.75,-1)
		(0.25,0.5) -- (0.75,0.5)
		(0.25,-0.5) -- (0.75,-0.5)
		(0.5,-0.45) -- (0.6,-0.55)
		(0.5,-0.95) -- (0.6,-1.05)
		(-0.5,0) -- (-0.25,0)
		(0.25,0) -- (0.75,0)
		;
		\node at (-1,0.75) {#1};
		\node at (-1,-0.75) {#2};
		\node at (1.2,0.75) {#3};
		\node at (-1,0) {#4};
		\node[meter, scale=1, very thin] (meter) at (1,0) {};
		\node at (1.75,0) {#5};
	\end{tikzpicture}
}

\newcommand{\distill}[3]{
	\begin{circuitikz}
		\node[rectangle,minimum width=0.5cm,minimum height=1cm,draw] at (0,0) {$U$};
		\draw [thick] (-0.5,0.25) -- (-0.25,0.25)
		(-0.5,-0.25) -- (-0.25,-0.25)
		(-0.5,0.25) -- (-0.5,0.75)
		(-0.5,0.75) -- (0.75,0.75)
		(0.25,0.25) -- (0.75,0.25)
		(0.25,-0.25) -- (0.75,-0.25);
		\node at (-1,0.5) {#1};
		\node at (-1,-0.25) {#2};
		\node at (1.5,-0.25) {#3};
		\node[meter, scale=1.5, very thin] (meter) at (1.25,0.5) {};
		\node[rectangle,minimum width=0.5cm,minimum height=1cm] at (2.1,0.25) {$P_k$};
	\end{circuitikz}
}

\newcommand{\distillblock}[5]{
	\begin{circuitikz}
		\node[rectangle,minimum width=0.5cm,minimum height=1.5cm,draw] at (0,-0.25) {$U$};
		\draw [thick] (-0.5,0.25) -- (-0.25,0.25)
		(-0.5,-0.75) -- (-0.25,-0.75)
		(-0.5,0.25) -- (-0.5,0.75)
		(-0.5,0.75) -- (0.75,0.75)
		(0.25,0.25) -- (0.75,0.25)
		(0.25,-0.75) -- (0.75,-0.75)
		(-0.5,-0.25) -- (-0.25,-0.25)
		(0.25,-0.25) -- (0.75,-0.25)
		;
		\node at (-1,0.5) {#1};
		\node at (-1,-0.75) {#2};
		\node at (1.5,-0.8) {#3};
		\node[meter, scale=1.5, very thin] (meter) at (1.25,0.5) {};
		\node[rectangle,minimum width=0.5cm,minimum height=1cm] at (2.1,0.25) {$P_m$};
		\node at (-1,-0.25) {#4};
		\node[meter, scale=1, very thin] (meter) at (1,-0.25) {};
		\node at (1.75,-0.25) {#5};
	\end{circuitikz}
}

\newcommand{\distillout}{
	\begin{circuitikz}
		\node[rectangle,minimum width=0.5cm,minimum height=0.5cm,draw] at (-0.15,0.25) {$A_k$};
		\node[rectangle,minimum width=0.5cm,minimum height=0.5cm,draw] at (-0.15,-0.75) {$B_k$};
		\draw [thick] (-0.75,0.25) -- (-0.5,0.25)
		(-0.75,-0.75) -- (-0.5,-0.75)
		(-0.75,0.25) -- (-0.75,0.75)
		(-0.75,0.75) -- (1.75,0.75)
		(0.2,0.25) -- (0.75,0.25)
		(0.2,-0.75) -- (0.75,-0.75)
		(1.75,0.75) -- (1.75,0.25)
		(1.75,0.25) -- (1.55,0.25)
		;
		\node[rectangle,minimum width=0.5cm,minimum height=0.5cm,draw] at (1.15,0.25) {$A_m^\dagger$};
		\node[circle, radius=1.5cm, draw] at (1.2,-0.75) {$\ket{\psi}$};
	\end{circuitikz}
}

\newcommand{\EmuTermOne}{
	\begin{circuitikz}
		\node[rectangle,minimum width=0.7cm,minimum height=0.5cm,draw] at (0,0.25) {$U$};
		\node[rectangle,minimum width=0.7cm,minimum height=0.5cm,draw] at (0,-0.75) {$U$};
		\node[rectangle,minimum width=0.7cm,minimum height=0.5cm,draw] at (2,0.25) {$U^\dagger$};
		\node[rectangle,minimum width=0.7cm,minimum height=0.5cm,draw] at (2,-0.75) {$U^\dagger$};
		\draw [thick] (-0.7,0.15) -- (-0.35,0.15)
		(-0.9,0.35) -- (-0.35,0.35)
		(-0.6,-0.85) -- (-0.35,-0.85)
		(-0.8,-0.65) -- (-0.35,-0.65)
		(0.35,0.15) -- (0.6,0.15)
		(0.35,0.35) -- (0.6,0.35)
		(0.35,-0.85) -- (0.6,-0.85)
		(0.35,-0.65) -- (0.6,-0.65)
		(2.6,-0.85) -- (2.35,-0.85)
		(2.7,-0.65) -- (2.35,-0.65)
		(2.7,0.15) -- (2.35,0.15)
		(2.6,0.35) -- (2.35,0.35)
		(1.65,-0.85) -- (1.4,-0.85)
		(1.65,-0.65) -- (1.4,-0.65)
		(1.65,0.15) -- (1.4,0.15)
		(1.65,0.35) -- (1.4,0.35)
		
		(0.6,0.35) -- (1.4,-0.65)
		(0.6,0.15) -- (1.4,0.15)
		(1.4,0.35) -- (0.6,-0.65)
		(0.6,-0.85) -- (1.4,-0.85)
		(-0.6,-0.85) -- (-0.6,-1.15)
		(2.6,-0.85) -- (2.6,-1.15)
		(-0.6,-1.15) -- (2.6,-1.15)
		(2.7,0.15) -- (2.7,0.65)
		(-0.7,0.15) -- (-0.7,0.65)
		(-0.7,0.65) -- (2.7,0.65)
		(2.6,0.35) -- (2.6,0.85)
		(-0.8,-0.65) -- (-0.8,0.85)
		(-0.8,0.85) -- (2.6,0.85)
		(2.7,-0.65) -- (2.7,-1.3)
		(-0.9,-1.3) -- (-0.9,0.35)
		(-0.9,-1.3) -- (2.7,-1.3)
		;
	\end{circuitikz}
}

\newcommand{\EmuTermOneAB}[3]{
	\begin{circuitikz}
		\node[rectangle,minimum width=0.7cm,minimum height=0.5cm,draw] at (0,0.25) {${#1}_k$};
		\node[rectangle,minimum width=0.7cm,minimum height=0.5cm,draw] at (0,-0.75) {${#1}_m$};
		\node[rectangle,minimum width=0.7cm,minimum height=0.5cm,draw] at (1.5,0.25) {${#1}_{#2}^\dagger$};
		\node[rectangle,minimum width=0.7cm,minimum height=0.5cm,draw] at (1.5,-0.75) {${#1}_{#3}^\dagger$};
		\draw [thick] 
		(-0.6,0.25) -- (-0.35,0.25)
		(0.35,0.25) -- (1.15,0.25)
		(1.85,0.25) -- (2.1,0.25)
		(2.1,0.25) -- (2.1,0.75)
		(-0.6,0.25) -- (-0.6, 0.75)
		(-0.6,0.75) -- (2.1, 0.75)
		
		(-0.6,-0.75) -- (-0.4,-0.75)
		(0.4,-0.75) -- (1.15,-0.75)
		(1.85,-0.75) -- (2.1,-0.75)
		(2.1,-0.75) -- (2.1,-1.25)
		(-0.6,-0.75) -- (-0.6, -1.25)
		(-0.6,-1.25) -- (2.1, -1.25)
		;
	\end{circuitikz}
}

\newcommand{\EmuTermTwo}{
	\begin{circuitikz}
		\node[rectangle,minimum width=0.7cm,minimum height=0.5cm,draw] at (0,0.25) {$U$};
		\node[rectangle,minimum width=0.7cm,minimum height=0.5cm,draw] at (0,-0.75) {$U$};
		\node[rectangle,minimum width=0.7cm,minimum height=0.5cm,draw] at (2,0.25) {$U^\dagger$};
		\node[rectangle,minimum width=0.7cm,minimum height=0.5cm,draw] at (2,-0.75) {$U^\dagger$};
		\draw [thick] (-0.7,0.15) -- (-0.35,0.15)
		(-0.9,0.35) -- (-0.35,0.35)
		(-0.6,-0.85) -- (-0.35,-0.85)
		(-0.8,-0.65) -- (-0.35,-0.65)
		(0.35,0.15) -- (0.6,0.15)
		(0.35,0.35) -- (0.6,0.35)
		(0.35,-0.85) -- (0.6,-0.85)
		(0.35,-0.65) -- (0.6,-0.65)
		(2.6,-0.85) -- (2.35,-0.85)
		(2.7,-0.65) -- (2.35,-0.65)
		(2.7,0.15) -- (2.35,0.15)
		(2.6,0.35) -- (2.35,0.35)
		(1.65,-0.85) -- (1.4,-0.85)
		(1.65,-0.65) -- (1.4,-0.65)
		(1.65,0.15) -- (1.4,0.15)
		(1.65,0.35) -- (1.4,0.35)
		
		(0.6,0.15) -- (1.4,-0.85)
		(0.6,0.35) -- (1.4,0.35)
		(1.4,0.15) -- (0.6,-0.85)
		(0.6,-0.65) -- (1.4,-0.65)
		(-0.6,-0.85) -- (-0.6,-1.15)
		(2.6,-0.85) -- (2.6,-1.15)
		(-0.6,-1.15) -- (2.6,-1.15)
		(2.7,0.15) -- (2.7,0.65)
		(-0.7,0.15) -- (-0.7,0.65)
		(-0.7,0.65) -- (2.7,0.65)
		(2.6,0.35) -- (2.6,0.85)
		(-0.8,-0.65) -- (-0.8,0.85)
		(-0.8,0.85) -- (2.6,0.85)
		(2.7,-0.65) -- (2.7,-1.3)
		(-0.9,-1.3) -- (-0.9,0.35)
		(-0.9,-1.3) -- (2.7,-1.3)
		;
	\end{circuitikz}
}

\newcommand{\EmuTermTwoA}[3]{
	\begin{circuitikz}
		\node[rectangle,minimum width=0.7cm,minimum height=0.5cm,draw] at (0,0.25) {${#1}_k$};
		\node[rectangle,minimum width=0.7cm,minimum height=0.5cm,draw] at (0,-0.75) {${#1}_m$};
		\node[rectangle,minimum width=0.7cm,minimum height=0.5cm,draw] at (1.5,0.25) {${#1}_{#2}^\dagger$};
		\node[rectangle,minimum width=0.7cm,minimum height=0.5cm,draw] at (1.5,-0.75) {${#1}_{#3}^\dagger$};
		\draw [thick] 
		(-0.6,0.25) -- (-0.35,0.25)
		(0.35,0.25) -- (1.15,0.25)
		(1.85,0.25) -- (2.1,0.25)
		(2.1,0.25) -- (2.1,0.75)
		(-0.7,-0.75) -- (-0.7, 0.75)
		(-0.7,0.75) -- (2.1, 0.75)
		
		(-0.7,-0.75) -- (-0.4,-0.75)
		(0.4,-0.75) -- (1.15,-0.75)
		(1.85,-0.75) -- (2.1,-0.75)
		(2.1,-0.75) -- (2.1,-0.25)
		(-0.6,-0.25) -- (-0.6, 0.25)
		(-0.6,-0.25) -- (2.1, -0.25)
		;
	\end{circuitikz}
}

\newcommand{\EmuTermTwoB}[3]{
	\begin{circuitikz}
		\node[rectangle,minimum width=0.7cm,minimum height=0.5cm,draw] at (0,0.25) {${#1}_k$};
		\node[rectangle,minimum width=0.7cm,minimum height=0.5cm,draw] at (0,-0.75) {${#1}_m$};
		\node[rectangle,minimum width=0.7cm,minimum height=0.5cm,draw] at (1.5,0.25) {${#1}_{#2}^\dagger$};
		\node[rectangle,minimum width=0.7cm,minimum height=0.5cm,draw] at (1.5,-0.75) {${#1}_{#3}^\dagger$};
		\draw [thick] 
		(-0.6,0.25) -- (-0.35,0.25)
		(0.35,0.25) -- (1.15,-0.75)
		(0.35,-0.75) -- (1.15,0.25)
		(1.85,0.25) -- (2.1,0.25)
		
		(-0.6,-0.75) -- (-0.4,-0.75)
		(1.85,-0.75) -- (2.1,-0.75)
		(2.1,-0.75) -- (2.1,-1.25)
		(-0.6,-0.75) -- (-0.6, -1.25)
		(-0.6,-1.25) -- (2.1, -1.25)
		(2.1,0.25) -- (2.1,.75)
		(-0.6,0.25) -- (-0.6, .75)
		(-0.6,.75) -- (2.1,.75)
		;
	\end{circuitikz}
}

\newcommand{\MPSI}{
	\begin{tikzpicture}
		\filldraw[fill=white] (0,0) circle (0.5) node {$A$};
		\filldraw[fill=white] (2,0) circle (0.5) node {$B$};
		\draw [thick]    (1.5,0) -- (0.5,0)
		(2,0.5) -- (2,1)
		(0,0.5) -- (0,1)
		(2,-0.5) -- (2,-1)
		(0,-0.5) -- (0,-1)
		;
		\node[] at (-0.5,0.75) {$d_A$}; 
		\node[] at (2.5,0.75) {$d_B$}; 
		\node[] at (-0.5,-0.75) {$d_A$}; 
		\node[] at (2.5,-0.75) {$d_B$}; 
		\node[] at (1,0.25) {$D$}; 
	\end{tikzpicture}
}

\newcommand{\MPSII}{
	\begin{tikzpicture}
		\filldraw[fill=white] (0,0) circle (0.5) node {$A^{(2)}$};
		\filldraw[fill=white] (-2,0) circle (0.5) node {$A^{(1)}$};
		\filldraw[fill=white] (4,0) circle (0.5) node {$A^{(M)}$};
		\draw [thick] (-1.5,0) -- (-0.5,0)
		(1.5,0) -- (0.5,0)
		(2.5,0) -- (3.5,0)
		(-2,0.5) -- (-2,1)
		(4,0.5) -- (4,1)
		(0,0.5) -- (0,1)
		(-2,-0.5) -- (-2,-1)
		(4,-0.5) -- (4,-1)
		(0,-0.5) -- (0,-1)
		;
		\node[] at (0.5,0.75) {$d_{A_1^{(2)}}$}; 
		\node[] at (-1.5,0.75) {$d_{A_1^{(1)}}$}; 
		\node[] at (4.5,0.75) {$d_{A_1^{(M)}}$}; 
		\node[] at (0.5,-0.75) {$d_{A_1^{(2)}}$}; 
		\node[] at (-1.5,-0.75) {$d_{A_1^{(1)}}$}; 
		\node[] at (4.5,-0.75) {$d_{A_1^{(M)}}$}; 
		\node[] at (1,0.25) {$D_2$}; 
		\node[] at (-1,0.25) {$D_1$}; 
		\node[] at (3,0.25) {$D_{M-1}$}; 
		\node[] at (2,0) {...}; 
	\end{tikzpicture}
}

\begin{document}
\title{Quantum Tensor Product Decomposition from Choi State Tomography}

\author{Refik Mansuroglu \orcidlink{0000-0001-7352-513X}}
\email[]{Refik.Mansuroglu@fau.de}
\affiliation{Department of Physics, Friedrich-Alexander Universität Erlangen-Nürnberg (FAU), Staudtstraße 7, 91058 Erlangen}

\author{Arsalan Adil
\orcidlink{0000-0001-9422-7609}}
\affiliation{Center for Quantum Mathematics \& Physics and Department of Physics \& Astronomy\\ UC Davis, One Shields Ave, Davis CA.}

\author{Michael J. Hartmann \orcidlink{0000-0002-8207-3806}}
\affiliation{Department of Physics, Friedrich-Alexander Universität Erlangen-Nürnberg (FAU), Staudtstraße 7, 91058 Erlangen}

\author{Zoë Holmes
\orcidlink{0000-0001-6841-4507}}
\affiliation{\'Ecole Polytechnique F\'ed\'erale de Lausanne, Lausanne, Switzerland}

\author{Andrew T. Sornborger
\orcidlink{0000-0001-8036-6624}}
\affiliation{Information Sciences, Los Alamos National Laboratory, Los Alamos, NM, USA.}

\date{\today}

\begin{abstract}
\noindent
    The Schmidt decomposition is the go-to tool for measuring bipartite entanglement of pure quantum states. Similarly, it is possible to study the entangling features of a quantum operation using its operator-Schmidt, or tensor product decomposition. While quantum technological implementations of the former are thoroughly studied, entangling properties on the operator level are harder to extract in the quantum computational framework because of the exponential nature of sample complexity. Here we present an algorithm for unbalanced partitions into a small subsystem and a large one (the environment) to compute the tensor product decomposition of a unitary whose effect on the small subsystem is captured in classical memory while the effect on the environment is accessible as a quantum resource. This quantum algorithm may be used to make predictions about operator non-locality, effective open quantum dynamics on a subsystem, as well as for finding low-rank approximations and low-depth compilations of quantum circuit unitaries. We demonstrate the method and its applications on a time-evolution unitary of an isotropic Heisenberg model in two dimensions.
\end{abstract}

\maketitle

\section{Introduction}
\begin{figure*}[t]
	\centering\includegraphics[width=\textwidth]{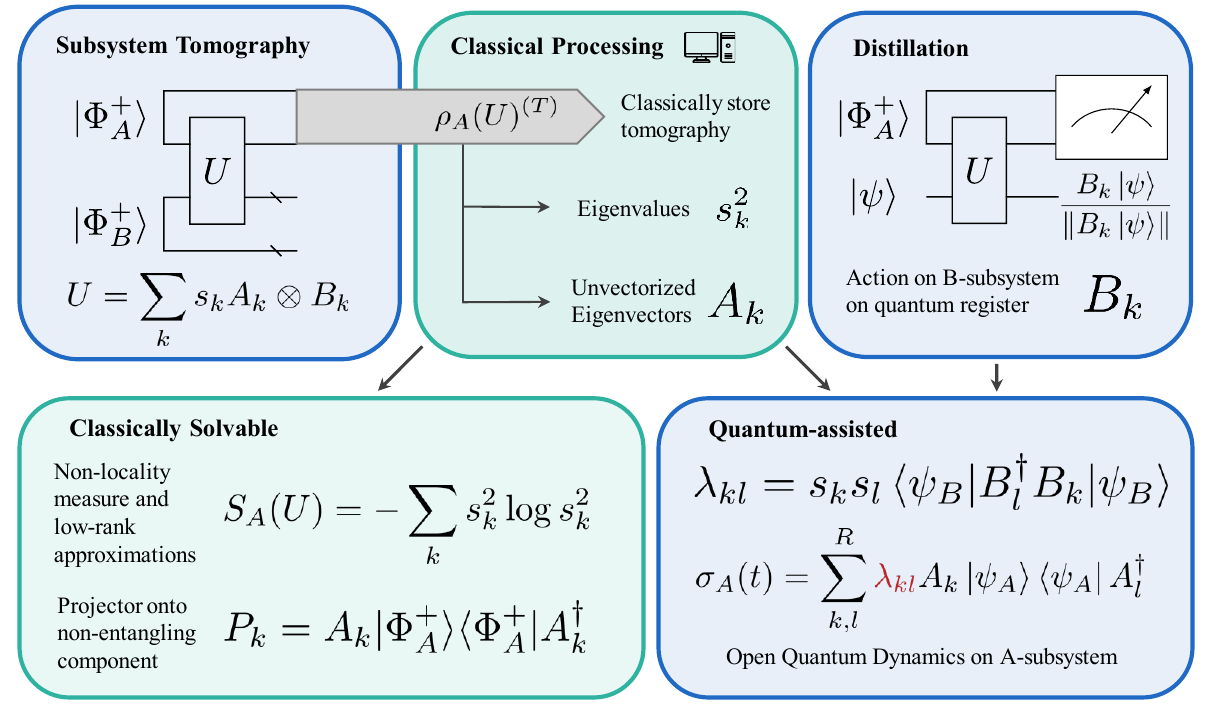}
	\caption{\textbf{Summary of QTPD.} Given a unitary operation $U$ as a quantum resource, it is decomposed into the form of Eq.~\eqref{eq:U_schmidt} in two steps. First, the Choi-state of $U$ is prepared and tomography is performed on the subsystem $\mathcal{H}_A$. The classical snapshot $\rho_T$ of the state $\rho$ of subsystem $\mathcal{H}_A$ is then classically diagonalized to obtain the tensors $A_k$ and the Schmidt values $s_k$ (cf. Eq.~\eqref{eq:snapshot}). With this classical information, the non-locality measure $S_A(U)$ introduced in Eq.~\eqref{eq:op_entanglement} can be calculated. The action of $U$ on the environment via $B_k$ is consequently obtained by a measurement of the observable $P_k$ from Eq.~\eqref{eq:dist} on the subsystem $\mathcal{H}_A$. The green boxes above denote fully classical steps, and the blue boxes denote steps where a quantum computer is used.}
	\label{fig:flow}
\end{figure*}
Entanglement is a defining feature of quantum theory \cite{Horodecki_2009}. The powerful capability of sharing information in a superposition of coupled states within a composite system still fascinates and puzzles physicists even a hundred years after the advent of quantum physics. Entanglement is used as a fundamental resource for quantum computing and has launched an entirely new paradigm for information processing \cite{arute_quantum_2019}.

For a fixed Hilbert space partition, $\mathcal{H} \cong \mathcal{H}_A \otimes \mathcal{H}_B$, the Schmidt decomposition of a pure state, $\ket{\psi}$, into tensor products, $\ket{\psi} = \sum_{k=1}^r \sigma_k \ket{a_k} \otimes \ket{b_k}$, reveals features about the shared entanglement between the two subsystems, $\mathcal{H}_{A}$ and $\mathcal{H}_{B}$. A disentangled state, or tensor product state, will consist of a single non-zero term, while an entangled state will have a Schmidt rank $r > 1$. Analogously, we can define the tensor product decomposition (TPD) \cite{VanLoan1993} or operator-Schmidt decomposition \cite{nielsen2000quantum} of a unitary operator, $U$, via
\begin{align}
	U = \sum_{k=1}^R s_k \, A_k \otimes B_k.
	\label{eq:U_schmidt}
\end{align}
Here,  $A_k \in \mathcal{L}(\mathcal{H}_{A}), B_k \in \mathcal{L}(\mathcal{H}_{B})$ are linear operators acting on the subsystems $\mathcal{H}_{A/B}$ and the rank, $R$, is the minimal number of non-zero terms in the TPD. Without loss of generality, we can impose the $A_k$ and $B_k$ to be orthogonal with respect to the Hilbert-Schmidt inner product, and normalized to $\norm{A_k}^2 = d_A \change{:= \dim(\mathcal{H}_A)}$ and $\norm{B_k}^2 = d_B \change{:= \dim(\mathcal{H}_B)}$ \change{using the 2-norm $\norm{.}$.} \change{Note that $A_k$ and $B_k$ are not unitary, in general.} Furthermore, the $s_k$ are non-negative, real numbers which are constrained to sum up to one, by unitarity of $U$, i.e. $\sum_k^R s_k^2 = 1$ (see App.~\ref{app:ambiguities} for details).

As a theoretical tool, TPD has previously been used to classify non-local and entangling content of unitaries \cite{D_r_2002, Jonnadula_2020, balakrishnan2011operatorschmidt}, and also for the analysis of time evolution of quantum many body systems \cite{Prosen_2007} and quantification of quantum chaos \cite{Zhou_2017, Bertini_2020}. Recently, TPD has been used to construct entanglement witnesses \cite{zhang2023analyzing}. While these advances motivate a systematic method to obtain the TPD of a quantum operator, current approaches are limited to classical resources. 

Van Loan and Pitsianis \cite{VanLoan1993, van2000ubiquitous} developed a classical algorithm to find the TPD of an operator $T$, not necessarily unitary, using the singular value decomposition of a reordered version of $T$. In quantum information processing, the operator of interest will typically require classical memory that grows exponentially in the number of qubits, making such classical methods inaccessible. While the measurement of Schmidt decompositions on quantum states has already been thoroughly studied \cite{smith2017quantifying, Subramanian_2021, zhang2023quantification}, works on the operator level remain limited to specific problems that can be treated analytically \cite{D_r_2002, balakrishnan2011operatorschmidt, Tyson_2003}. 

Here we bring the tensor product decomposition into a quantum algorithmic framework. In particular, we present a hybrid quantum-classical algorithm that performs the quantum tensor product decomposition (QTPD) described in Eq.~\eqref{eq:U_schmidt} for a unitary matrix $U$ with a known quantum circuit representation. If we assume an asymmetric split for which \change{$d_A \ll d_B$}, QTPD provides the operators $A_k$ in classical memory, whereas $B_k$ are accessed as a quantum resource distilled out of $U$. The complete algorithm is visualized in Fig.~\ref{fig:flow}.

We discuss a number of immediate applications and new directions for future research that are enabled with QTPD. Alongside low-rank approximations, QTPD provides a tool for studying entanglement, with application to entanglement witnesses and measures of entanglement generation \cite{Jonnadula_2020, zhang2023analyzing}, classically assisted simulation of (open) quantum dynamics \cite{Mansuroglu_2023, mansuroglu2023problem, Mc_Keever_2023} and low-depth compilation techniques \cite{meyer2023quantum}. The fact that the $A_k$ are stored classically goes hand in hand with the philosophy of hybrid quantum computing, which is to use quantum resources as little as possible, but in the most crucial step. We note, however, that QTPD collects all necessary data from the quantum computer at the start of the algorithm, and does not require a hybrid quantum-classical optimization loop~\cite{cerezo2023does}. We demonstrate QTPD and its applications on the time evolution operator of an isotropic Heisenberg model.

\section{Quantum Tensor Product Decomposition}
\change{
	We introduce QTPD in two steps. First, a matrix representation of the action on $\mathcal{H}_A$ is captured via tomography and second, we discuss a distillation technique to subsequently capture the action on $\mathcal{H}_B$ as a quantum channel. We finally compare required resources for QTPD against its classical competitor, discuss possible adaptations for the near term and comment on error propagation.
}

\subsection{The Algorithm}
We start with a unitary operator $U$ that is accessible as a quantum resource (i.e. it is accessible as an oracle or its circuit representation is known). Our aim is to get a classical snapshot of the reduced action of $U$ on $\mathcal{H}_A$, which is implicit in the operators $A_k$ in Eq.~\eqref{eq:U_schmidt}. Consider the action of $U$ on the two generalized Bell states 
\begin{align}
	\ket{\Phi^+_{A/B}}& = \frac{1}{\sqrt{d_{A/B}}} \sum_{i=1}^{d_{A/B}} \ket{i} \ket{i} \; ,
\end{align}
that are states on two copies of the subsystems $\mathcal{H}_A$ and $\mathcal{H}_B$, respectively (cf. first circuit in Fig.~\ref{fig:flow}). After tracing out $\mathcal{H}_B$ from $U \left(\ket{\Phi^+_A} \otimes \ket{\Phi^+_B}\right)$, we are left with the mixed state of the vectorizations $\mathrm{\mathrm{vec}}(A_k) = \frac{1}{\sqrt{d_A}} \sum_i^{d_A} \ket{i} A_k \ket{i}$, i.e.
\begin{align}
	\rho_A(U) = \frac{1}{d_A} &\sum_k^R s_k^2 \; \mathrm{\mathrm{vec}}(A_k) \mathrm{\mathrm{vec}}(A_k)^\dagger \; ,
	\label{eq:snapshot}
\end{align}
which can be derived using orthonormality of the $A_k$ and $B_k$ (see App.~\ref{app:snapshot}). Unvectorizing the $\mathrm{vec}(A_k)$ is exponentially hard, in general \cite{Sedl_k_2019}. Since $d_A$ is assumed to be much smaller than $d_B$, a tomography of the state $\rho_A(U)$ can be taken and stored as a classical snapshot. Diagonalizing $\rho_A(U)$ finally yields the eigenvectors $\mathrm{vec}(A_k)$ with corresponding eigenvalues $s_k^2$. \change{Note that this is mathematically equivalent to finding one half (on $\mathcal{H}_A$) of the Schmidt decomposition of the Choi state of $U$.}

The above algorithm not only yields information about $s_k$ and $A_k$, but we can also find $B_k$ as a quantum resource. Once the $A_k$ are found, one can distill out the individual $B_k$. That is, given a state $\ket{\psi}$ one can prepare $B_k \ket{\psi}$. This is done via a partial measurement of the projector
\begin{align}
	P_k = A_k \ket{\Phi^+_A} \bra{\Phi^+_A} A_k^\dagger  \label{eq:dist}
\end{align}
on the state $U \left( \ket{\Phi_A^+} \otimes \ket{\psi} \right)$ (cf. second circuit in Fig.~\ref{fig:flow} and see App.~\ref{app:distillation} for a derivation). \changeTwo{The projectors, $P_k$, are orthogonal, i.e. $P_k P_l = \delta_{kl} P_k$, which is a direct consequence of the orthonormality of the $A_k$. As a result, they can be simultaneously measured, such that every shot of the distillation circuit (cf. Fig.~\ref{fig:flow}) yields the normalized output state $\frac{B_k \ket{\psi}}{\norm{B_k \ket{\psi}}}$ with probability $p_k = s_k^2 \norm{B_k \ket{\psi}}^2$. Note that $\sum_{k=1}^R p_k = 1$ from unitarity of $U$, see App.~\ref{app:distillation} for details.
	
	If one is interested in the action of a specific $B_k$ only, the distillation process involves post-selection on the outcome of the partial measurement. This comes with a sample overhead of $\mathcal{O}\left( \frac{1}{p_k} \right) = \mathcal{O}\left( \frac{1}{s_k^2 \norm{B_k \ket{\psi}}^2 } \right)$, which is never a serious issue. The overhead becomes large when either $\norm{B_k \ket{\psi}}$ or $s_k$ become small. In the first case, the output state is close to the zero vector and in the second case, the sampled tensor component is a small contribution in a low-rank approximation of $U$.
}

QTPD can be used to determine \changeTwo{such} approximations to $U$ of a specified rank. A set of operators $\{C_k\}$ and $\{D_k\}$, such that the 2-norm
\begin{align}
	\norm{U - \sum_{k}^r \change{t_k} C_k \otimes D_k},
	\label{eq:low_rank}
\end{align}
is minimal, is called a rank $r$ approximation. \change{We introduced the positive, real-valued scalars $t_k$ following the same convention as in the tensor product decomposition of $U$.} The special case $r=1$ \change{from Eq.~\eqref{eq:low_rank}} corresponds to the well-known nearest Kronecker problem \cite{VanLoan1993, van2000ubiquitous}. The solution to minimize Eq.~\eqref{eq:low_rank} is the sum of product operators, $\sum_k^r \change{s_k} A_k \otimes B_k$, that correspond to the largest eigenvalues $\{s_k^2\}_{k=1}^r$ of $\rho_A(U)$ (cf. Eq.~\eqref{eq:snapshot}, see App.~\ref{app:lowrank} for a proof).

As the $s_k$ and the $A_k$ are classically stored, but the $B_k$ are not, we cannot classically store a low rank approximation. A low rank approximation can be used to suppress the sample complexity of QTPD whenever the sample budget is limited. This allows us to resolve just the singular values, $s_k$, that are sufficiently large and still provide a good approximation of $U$. In particular, if $\varepsilon$ is the tolerable error of resolving the largest eigenvalues of $\rho_A(U)$, then the sample complexity scales as \change{$\mathcal{O}\left( R \frac{d_A^2}{\varepsilon^2} \right)$} \cite{Apeldoorn23}. Hence, we achieve an $\varepsilon$-close approximation to $\rho_A(U)$ in the operator norm by dropping every $s_k^2 < \epsilon$. \change{A low-rank approximation, in which the $B_k$ are accessible on a quantum computer, as described above, can be seen as an application of QTPD. It differs from the applications that are discussed in section \ref{sec:applications}, as it is always applied along with tomography.}

\subsection{Resources for QTPD}
\change{
	QTPD has the obvious advantage over classical methods that the unitary $U$ can be loaded as a quantum circuit. Since the Bell pair creation can be executed in depth 2, the depth of QTPD is primarily given by the depth of the circuit $U$. The runtime of QTPD is hence dominated by the sample complexity of tomography that is in this case $\tilde{\mathcal{O}}\left( R \frac{d_A^2}{\epsilon^2} \right)$ \cite{Apeldoorn23}. In memory, QTPD comes with a linear overhead of $n$ ancilla qubits and requires classical memory to store $\mathcal{O}\left( R d_A^2 \right)$ complex numbers in the worst case.
	
	The B-distillation step is similar in depth and admits a smaller sample complexity $\mathcal{O}\left( s_{k_0}^{-2} \right)$ with $s_{k_0}$ the smallest coefficient to be resolved. It also only needs $n_A$ ancilla qubits and no additional classical memory.
	
	A classical method to find the TPD of $U$ goes back to van Loan and Pitsianis \cite{VanLoan1993}, who showed that the operator $\mathcal{R}(U) \in \mathrm{Mat}(d_A^2\times d_B^2, \mathds{C})$ with permuted elements, such that 
	\begin{align}
		\mathcal{R}(U) = \sum_k s_k \mathrm{vec}(A_k) \mathrm{vec}(B_k)^\dagger,
	\end{align}
	encodes the tensor factors in its singular value decomposition. The tensor factors $A_k$ and $B_k$ can be derived as the unvectorized left- and right-eigenvectors of $\mathcal{R}(U)$ and the $s_k$ are its singular values. If we neglect the runtime for reshaping $U$ into $\mathcal{R}(U)$, the bottleneck of van Loan and Pitsianis' algorithm comes from the numerical solution for the singular value decomposition, which is $\mathcal{O}\left( d_B^2 d_A^2 R \right)$ \cite{trefethen1997numerical, Halko11}. The factor $R$ can be further weakened by randomization techniques \cite{Halko11}.
	
	This together with the required classical memory to store the $d^2$ complex elements of $U$ make the classical algorithm uncompetitive for large dimensions $d$. To be precise, QTPD achieves a superpolynomial speedup of a factor $\tilde{\mathcal{O}}\left( d_B^2 \epsilon^2 \right)$ and memory savings by a factor of $\frac{d_B^2}{R}$, which both grow exponentially in system size. If $U$ is given as a sparse matrix, Krylov subspace methods can be employed, which lower the required memory, but not the runtime \cite{Halko11}.
}

\subsection{Circumvention of Doubling the System Size}
\change{
	Next to the inevitable scaling in subspace dimension \cite{Sedl_k_2019}, the largest challenge within QTPD in the near term will be to keep the entangled Bell pairs coherent until measurement. Further, small quantum processors with fast read-out and long coherence times are most efficient when memory requirements are traded off against runtime.
	
	To this end, the effect of the Bell pairs $\ket{\Phi^+_{A/B}}$ is reduced to an average over a basis of $\mathcal{H}_A$ and $\mathcal{H}_B$, respectively. In every run, a random initial state drawn from a basis of choice (for instance the computational basis) is fixed. The output state for a fixed basis state $\ket{j_A}$ of $\mathcal{H}_A$ and $\ket{j_B}$ of $\mathcal{H}_B$ is
	\begin{align}
		\sum_k s_k A_k \ket{j_A} \otimes B_k \ket{j_B}.
	\end{align}
	Averaging over the basis in $\mathcal{H}_B$ accounts for an effective partial trace, using
	\begin{align}
		\mathds{E}_{\ket{b}}[\braket{b | B_k B_l | b}] &= \frac{1}{d_B} \sum_{j = 1}^{d_B} \braket{j | B_k B_l^\dagger | j} \nonumber \\
		&= \frac{1}{d_B} \Tr( B_k B_l^\dagger ) = \delta_{kl},
	\end{align}
	where the expectation value is taken over the discrete set of basis states in $\mathcal{H}_B$. After tracing out $\mathcal{H}_B$, we are thus left with
	\begin{align}
		\sum_k s_k^2 A_k \ket{j_A} \bra{j_A} A^\dagger_k.
		\label{eq:fixed_state}
	\end{align}
	If we keep the input-output relation within $\mathcal{H}_A$, the vectorization of the $A_k$ can be reconstructed using tomography and summing over all basis states in $\mathcal{H}_A$ 
	\begin{align}
		\sum_{j=1}^{d_A} \ket{j} A_k \ket{j} = \sqrt{d_A} \mathrm{vec}(A_k).
	\end{align}
	Recall the definition of the normalized vectorization above Eq.~\eqref{eq:snapshot}. As a result, the same state as in Eq.~\eqref{eq:snapshot}, coming from the parallelized version using the Choi-state, can be reconstructed in classical post-processing. As opposed to the Choi-state based version, however, the runtime of this sequential version of QTPD is increased by the repeated tomography for a fixed basis state $\ket{j} \in \mathcal{H}_A$, yielding an overhead factor of $d_A$ in runtime. Convergence to the mean value from averaging over $\mathcal{H}_B$, on the other hand, does not introduce an additional sample overhead, as it is equivalent to tracing out $\mathcal{H}_B$ as part of the Choi-state-based approach.
}

\subsection{Error Analysis}
\change{
	In general, the error from shot noise in tomography will be operator-valued and can be viewed as the difference between the correct state, $\rho_A(U)$, and the classical snapshot, $\rho_A(U)^{(T)}$, i.e. $\epsilon^{(T)} = \rho_A(U) - \rho_A(U)^{(T)}$. The shot noise error, $\epsilon^{(T)}$, propagates through QTPD and thus introduces errors to the $s_k, A_k$ and $B_k$. We discuss the technical details in App.~\ref{app:errors_QTPD} and briefly present the results here. In order to understand how errors propagate, we consider the errors of the eigenvalues and eigenvectors of $\rho_A(U)^{(T)}$ separately
	\begin{align}
		\rho_A(U) &= \rho_A(U)^{(T)} + \epsilon^{(T)} \nonumber \\
		&= (V - \epsilon^{(V)}) (D - \epsilon^{(D)}) (V - \epsilon^{(V)})^\dagger + \epsilon^{(T)}.
	\end{align}
	Our aim is to express the errors of $s_k$ and $A_k$ via $\epsilon = \max \left( \norm{\epsilon^{(D)}}, \norm{\epsilon^{(V)}} \right)$. To this end, we define the error measures $\epsilon^{(S)}_k = s_k^2 - {s_k^{(T)}}^2$ and $\epsilon_k^{(A)} = A_k - A_k^{(T)}$. With these measures, we can trivially relate
	\begin{align}
		\norm{\epsilon^{(D)}} &= \norm{D - D^{(T)}} = \sqrt{\sum_k \left( \epsilon^{(S)}_k \right)^2} \leq d_A \epsilon^{(S)},
	\end{align}
	where the factor $d_A$ can be removed by appropriate normalization of the 2-norm in order to represent an average case error. Less trivially, but after a straightforward calculation (see App.~\ref{app:errors_QTPD}), we can also relate 
	\begin{align}
		\norm{\epsilon^{(A)}_{k}} &= \norm{A_k - A_k^{(T)}} = \sqrt{ - 2 \left| \braket{A_k^{(T)} | \epsilon_k^{(A)}} \right| },
	\end{align}
	and with this finally
	\begin{align}
		\norm{\epsilon^{(V)}} &= \norm{V - V^{(T)}} = \sqrt{ - 2 \sum_{k,l} \left| \frac{\braket{A_k^{(T)} | \epsilon_l^{(A)}}}{d_A} \right| }.
	\end{align}
	We can show that $\norm{\epsilon^{(T)}} \leq 3 \epsilon$, which completes the propagation from tomography error to the tensor factors $A_k$. Applying the faulty $A_k^{(T)}$ for distillation of the effective action $B_k^{(T)}$ as a quantum resource imposes further error propagation on the projector $P_k$, defined in Eq.~\eqref{eq:dist}. Taking into account appropriate (faulty) normalization factors, we can also bound the error of $B_k^{(T)}$ in the following way
	\begin{align}
		&\norm{ \left( B_k - B_k^{(T)} \right) \ket{\psi} } \leq \left( 1+\frac{1}{\sqrt{2}}\right) \frac{1}{\sqrt{d_A}} \norm{\epsilon_k^{(A)}} + \mathcal{O}(\epsilon^2).
	\end{align}
	Altogether we show that the error contributions in $s_k$, $A_k$ and $B_k$ are linear in the 2-norm $\norm{\epsilon^{(T)}}$.
}

\section{Applications}
\label{sec:applications}
The quantum tensor product decomposition allows us to find and store the tensor components $A_k$ in classical resources and $B_k$ in a quantum resource. With this, we can solve a number tasks of interest. 

\subsection{Non-locality}
One application of QTPD lies in measuring the non-locality of the action of $U$, also called operator entanglement entropy \cite{Prosen_2007, Zhou_2017, Bertini_2020}, which further bounds how much entanglement $U$ generates. The vectorization of $U$ allows for a mapping of operators to quantum states. On this space, we can employ entanglement entropy measures that are defined for states. Consider the vectorized operator
\begin{align}
	\mathrm{vec}(U) = \sum_k^R s_k \, \mathrm{vec}(A_k) \otimes \mathrm{vec}(B_k).
\end{align}
If we trace out the $B$-subsystem, we get exactly the mixed state of Eq.~\eqref{eq:snapshot}. The von Neumann entanglement entropy of this state reads
\begin{align}
	S_A(U) = - \sum_k^R s_k^2 \log s_k^2 ,
	\label{eq:op_entanglement}
\end{align}
and is a measure of the non-locality of the action of $U$. The non-locality $S_A(U)$, sometimes referred to as the Schmidt strength \cite{balakrishnan2011operatorschmidt, Tyson_2003}, can be determined classically after a successful QTPD and admits a linear contribution from the tomography error to leading order (cf. App.~\ref{app:errors_applications}). 

Note that, although the non-locality of product operations vanishes, $S_A(A \otimes B) = 0$, Eq.~\eqref{eq:op_entanglement} alone is not a good measure of entanglement generation. For instance the swap gate, which maps product states to product states, reaches the maximal value for Eq.~\eqref{eq:op_entanglement}. To measure entanglement generation, one considers the entangling power of a circuit \cite{Jonnadula_2020, Zanardi_2000, Eisert_2021}. Several measures for entangling power have been proposed, of which we discuss two in App.~\ref{app:entanglement_gen}.

\subsection{Mereology}
If $U$ is generated by a physical Hamiltonian, one might be interested in searching for a bipartite factorization (sometimes referred to as the tensor product structure) of the global Hilbert space such that the two subsytems are decoupled, i.e. $U = U_A \otimes U_B$. Concretely, consider a Hamiltonian describing two interacting subsystems, $H = \sum_i H^A_i \otimes H^B_i $, where $H^A_i$ and $H^{B}_i$ are operators acting on states describing subsystems $A$ and $B$ in a Hilbert space factorized as $\mathcal{H} = \mathcal{H}_A\otimes \mathcal{H}_B$. Since this factorization is essentially a particular choice of a global basis, it can be related to another one by a unitary \cite{carroll2021quantum}, $\mathcal{H}_A \otimes \mathcal{H}_B \xrightarrow{V} \mathcal{H}_{A'} \otimes \mathcal{H}_{B'}$, where $V$ is a (non-local) unitary and states  in $A'$ and $B'$ describe physically different subsystems than $A$ and $B$. In particular, it may be possible to find a factorization of $\mathcal{H}$ such that the Hamiltonian is decoupled, i.e. $VHV^\dagger = H_{A'} + H_{B'}$. 

Some have used this approach, with the goal of minimizing the interaction Hamiltonian between two subsystems, to understand the emergence of classicality \cite{tegmark2015consciousness}. Practically, this approach appears in cases where taking a certain transformation can lead to analytically tractable forms of the Hamiltonian, e.g. in the case of the Jordan-Wigner transformation that  transforms certain interacting qubit Hamiltonians to a set of free fermionic operators. While QTPD does not itself find the optimal basis $V$ that will lead to approximately decoupled dynamics, it can be used to evaluate the cost function as part of another algorithm (such as the one proposed in \cite{adil2024search}). Two candidates to minimize are $\sum_{k=2}^R s_k$ or $1 - s_1^2$ where the $s_k$ are the singular values in the tensor product decomposition of the time propagator $U = Ve^{-iHt}V^\dagger = \sum_k^R s_k A_k \otimes B_k$.
The existence of such a \textit{decoherence-free} split \cite{Lidar_2003, Mansuroglu_2023_nop, Bernards_2024} is tightly connected to spectral properties of $U$, see App.~\ref{app:entanglementgrowth} for an example and App.~\ref{app:decfreesub} for a necessary and sufficient condition.

\subsection{Fast Quantum Transform and Classical Simulability}
A mereology algorithm can be further utilized to find an efficient compilation of a target unitary $U$. If there exists a basis in which $U$ decouples, $U = V^\dagger (U_{A} \otimes U_B) V$ can be implemented with a single layer after rotating into the basis $V$. A divide-and-conquer approach successively reduces the action of $U$ into a tensor product of $M$ local gates, i.e. $VUV^\dagger = U_{A^{(1)}} \otimes U_{A^{(2)}} \otimes ... \otimes U_{A^{(M)}}$. Such a fast quantum transform requires a rotation into the basis $V$, which is entangling, in general. 

More generally, for an arbitrary $U$, the closest fast quantum transform can be found via iterative QTPD, which can be performed efficiently if there is a single dominant coefficient $s_{1...1}$ in the multi-partite factorization 
\begin{align}
	U = \sum_{j_1,...,j_M}^{R_1, ..., R_M} s_{j_1 ... j_M} A_{j_1}^{(1)} \otimes ... \otimes A_{j_M}^{(M)}.
	\label{eq:multi}
\end{align}
We used the letter $A$ for all operators to emphasize that the local dimensions are small enough to be classically simulated. 

We can use the nearest unitary representations $U_m$ of the tensor factors $A_1^{(m)}$ of the rank one approximation of $U$ to construct a fast quantum transform approximation to $U$. 
\change{
	We show that the error is $\mathcal{O}(\sqrt{1-s_{1...1}})$. For this, we separate the error of the fast quantum transform into two terms. One representing the error from truncating all terms from $U$ except $s_{1...1}$ and one capturing the error by nearest unitary approximation $\frac{1}{ \sqrt{2 d_{A^{(m)}}} } \norm{A_1^{(m)} - U_m} = \epsilon^{(m)} > 0$. Writing $1-s_{1...1} = \epsilon_s > 0$, we get
	\begin{align}
		\frac{1}{ \sqrt{2 d} } \norm{U - U_1 \otimes ... \otimes U_M}& \nonumber \\
		\leq \sqrt{ \epsilon_s } + &\sqrt{ \frac{1}{2} \epsilon_s^2 + \sum_{m=1}^M \left( \epsilon^{(m)} \right)^2 },
	\end{align}
	see App.~\ref{app:nearestUnitary} for details.}
\begin{figure}
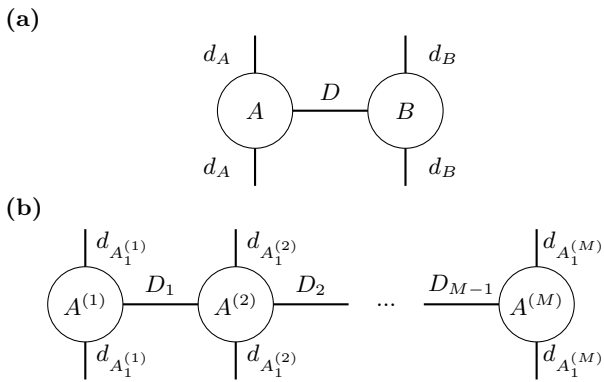

	\raggedright \textbf{(a)} \\
	\centering
	\MPSI
	
	\raggedright\textbf{(b)} \\ 
	\centering \MPSII
	\caption{\textbf{Low-entanglement clustering for a matrix product state representation. (a)} Dividing into subsystems $A$ and $B$, a matrix product \change{operator representing $U$} requires a certain bond dimension $D \leq \change{d_A^2}$, which is upper bounded by the lower dimension \change{$d_A^2 \leq d_B^2$} and dependent on the entanglement between $A$ and $B$. \textbf{(b)} Multi-partite decomposition of $U$ into $A_{j_m}^{(m)}$ allows for estimating the necessary bond dimensions $D_m$, which can be deduced by a low rank approximation of the multipartite decomposition (cf. Eq.~\eqref{eq:multi}). }
	\label{fig:MPO}
\end{figure}
If there is not only one, but a polynomial number of dominant coefficients, a better approximation to the action of $U$ can be achieved through a rank $r$ approximation. In this case, the probability to sample from the dominant $B_k$ is suppressed by a polynomial factor. The simulation of a fast quantum transform, $U_{A^{(1)}} \otimes U_{A^{(2)}} \otimes ... \otimes U_{A^{(M)}}$, is not only classically efficient, it is also made up of low-entangling transformations. Instead of achieving a close unitary approximation, the goal here is to bound the bond dimensions necessary for a faithful tensor network representation.

Since the non-locality measure $S_A(U)$ bounds the entanglement generation, it can be used as a witness to scan for clusterings of the Hilbert space with low entanglement between the clusters. If a cluster $\mathcal{H}_{A^{(1)}}$, on which the action of $U$ is low-entangling, is found, the entanglement generation between a second cluster $\mathcal{H}_{A^{(2)}}$ and its environment $\mathcal{H}_{A^{(3)}} \otimes ... \mathcal{H}_{A^{(M)}}$ will be bounded (cf. Fig.~\ref{fig:MPO}), as well. If the non-locality of the full unitary $U$ is bounded, it can thus be written as a matrix product operator \cite{Pirvu_2010}, of which the fast quantum transform is an extremal case. This allows for an efficient classical representation of the output of $U$, for instance via matrix product states \cite{perezgarcia2007matrix} or projected entangled pair states \cite{Verstraete_2008}. 

Fast quantum transforms have conceptual similarities to entanglement forging \cite{Eddins_2022}, which is used to simulate a larger system by simulating the subsystems separately on a smaller quantum chip, if there are only few connecting gates in the compilation of $U$. As opposed to QTPD, these methods are typically concerned with symmetric splittings and aim for a reduction of quantum resources in the simulation of $U$ instead of finding classically simulable subsystems.

\subsection{Open Quantum dynamics}
QTPD is applicable to studying entangling dynamics, or the decoherence of subsystem A into subsystem B. If we start with a product state $\ket{\psi_A} \otimes \ket{\psi_B}$, the evolved state within subsystem A will be mixed. The effective open quantum dynamics can be written in the form
\begin{align}
	\sigma_A &= \Tr_B\left( U \ket{\psi_A} \otimes \ket{\psi_B} \bra{\psi_A} \otimes \bra{\psi_B} U^\dagger \right) \nonumber \\
	&= \sum_{k,l}^R \lambda_{kl} A_k \ket{\psi_A}\bra{\psi_A}A_l^\dagger
	\label{eq:open_dynamics}
\end{align}
with $\lambda_{kl} = s_k s_l \bra{\psi_B} B_l^\dagger B_k \ket{\psi_B}$. While the operators $A_k$ and the state $\ket{\psi_A}$ can be stored on a classical machine, the $\lambda_{kl}$ are not accessible, a priori. Instead, the overlaps $\bra{\psi_B} B_l^\dagger B_k \ket{\psi_B}$ have to be determined using modified Hadamard tests or swap tests \cite{modhad, nielsen2000quantum} with different outputs of the $B_k$ distillation via projective measurement of Eq.~\eqref{eq:dist}. 

Once the $\lambda_{kl}$ and $A_k$ are stored classically, it is possible to simulate the open dynamics via Eq.~\ref{eq:open_dynamics} for any initial state $\ket{\psi_A}$. Eq.~\eqref{eq:open_dynamics} can be transformed into its Kraus representation, for instance by diagonalizing the Choi matrix. In this manner, QTPD can be used as a quantum-enhanced classical simulation algorithm~\cite{cerezo2023does} for open system simulation. That is, quantum hardware is crucial to obtaining the $A_k$ but then Eq.~\ref{eq:open_dynamics} acts as a classical surrogate to simulate the dynamics of any initial state and observable. \change{Note the parallels to process learning \cite{caro2022learning, caro2022outofdistribution}, which aim for learning a quantum channel from measurements of local observables using classical resources, such as neural networks.}

The error in predicting observables on $\sigma_A$ can be bounded by the trace norm to the faulty prediction $\sigma_A^{(T)}$ from tomography, which scales linearly with the tomography error $\norm{\epsilon^{(T)}}$ as we show in App.~\ref{app:errors_applications}. A na\"ive quantum simulation with fixed initial state and fixed observable suffers from shot noise that has the same scaling in samples as $\epsilon^{(T)}$.

\begin{figure}
	\centering
	\includegraphics[scale=0.35]{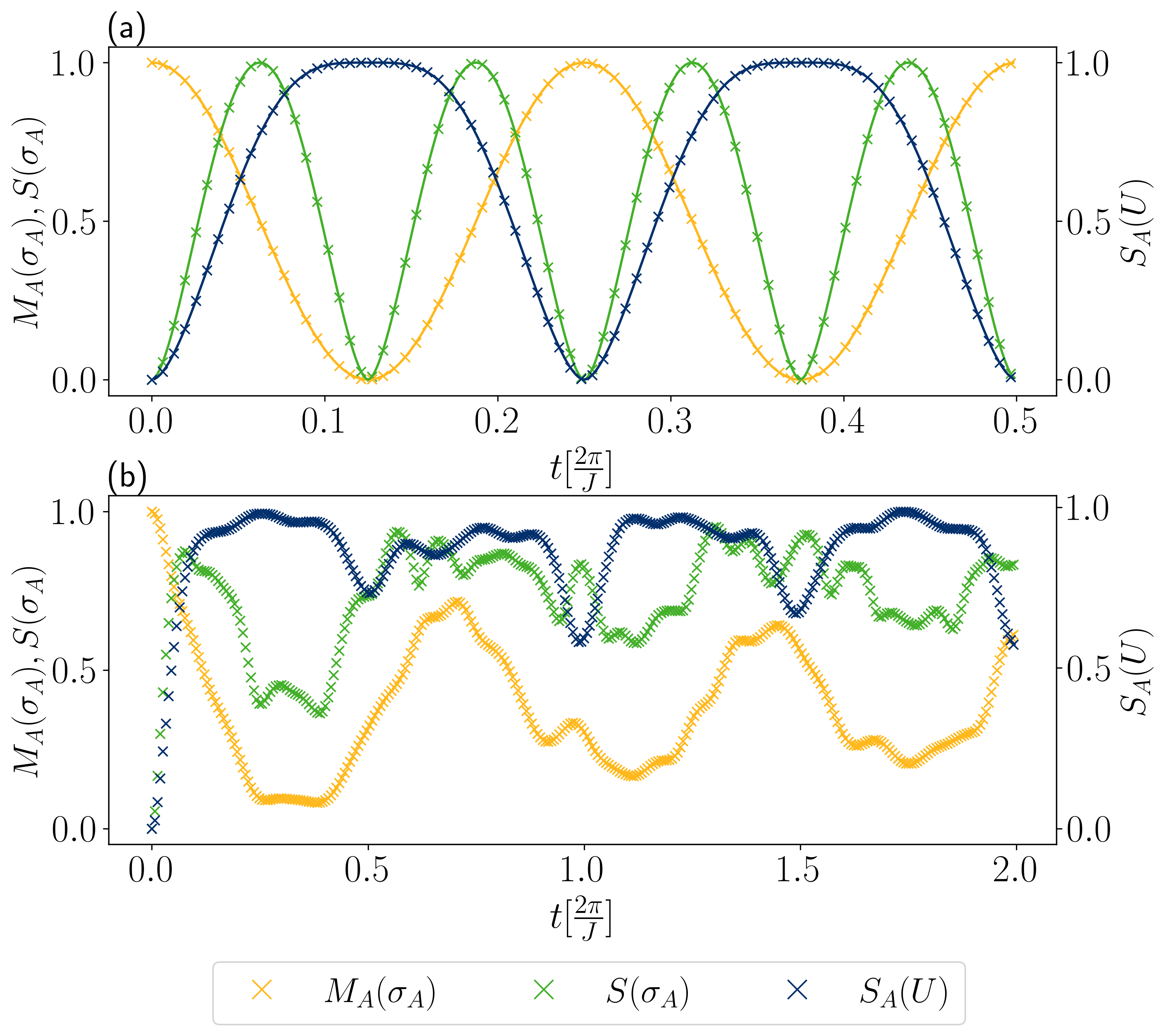}
	\caption{Open quantum dynamics for the isotropic Heisenberg model. We study a subsystem of one of two neighboring qubits (a) and two qubits of a 2D grid of $3\times 2$ qubits (b). The blue points show the non-locality measure $S_A(U)$ for the respective time evolution operator, while yellow and green contain the total magnetization $M_A(\sigma_A)$ and the entanglement entropy $S(\sigma_A)$ of the time evolved state. The initial state of the total system reads $\ket{1}\otimes \ket{0}$ in (a) and $\ket{11}\otimes \ket{0000}$ in (b). In (a), solid curves represent analytical predictions derived in App.~\ref{sec:Heisenberg}.}
	\label{fig:iso_Heisenberg}
\end{figure}

\section{Numerical Experiment}
We demonstrate QTPD on a Hamiltonian simulation problem for the isotropic Heisenberg model. To this end, we numerically solve the tensor decomposition by exact diagonalization of the unitary time evolution generated by the Hamiltonian
\begin{align}
	H = - J \sum_{\langle i,j \rangle} \left( X_i X_j + Y_i Y_j + Z_i Z_j \right)
	\label{eq:Heisenberg_nq}
\end{align}
with the Pauli matrices $\{X,Y,Z\}$ and the sum over nearest neighbors denoted by $\langle i,j \rangle$. We discuss one system of toy size whose dynamics we can analytically solve and a separate system on a two-dimensional grid that is small enough to be checked by exact diagonalization.

Let us consider a model of two qubits first. We show in App.~\ref{sec:Heisenberg} that the time evolution operator generated by the Heisenberg Hamiltonian incorporates an oscillation between the identity and the swap operator for certain times. In between those times, entanglement is alternately generated and reduced. It is thus natural to view each qubit as a subsystem and study the open dynamics on one of the two qubits. Larger systems can be split in two and the swapping of excitations from $\mathcal{H}_A$ to $\mathcal{H}_B$ and vice versa can be studied, as well. 

When more than two qubits are exchanging excitations, the overall dynamics are an ensemble of interfering oscillations, which depend on the initial state and the geometry of the interaction graph. To study both the distribution of excitations, and the entanglement between the split into subsystems $\mathcal{H}_A$ and $\mathcal{H}_B$, we stroboscopically measure the total magnetization $M(\sigma_A(t))$ of subsystem $\mathcal{H}_A$, as well as the von-Neumann entanglement entropy of the output state $S(\sigma_A(t))$
\begin{align}
	M_A(\sigma_A(t)) &= \Tr\left(\sum_{i \in \mathcal{I}_A} Z_i \sigma_A(t)\right) \label{eq:MA}
	\\
	S(\sigma_A(t)) &= -\Tr\left(\sigma_A(t) \log\sigma_A(t) \right), \label{eq:S}
\end{align}
with $\mathcal{I}_A$ denoting the index set for qubits in subsystem $\mathcal{H}_A$. For the sake of a clear comparison, we normalize all observables to take values between 0 and 1. This means we divide state entanglement entropies by $\log(d)$ and operator non-localities by $\log(d^2)$ with $d$ being the dimension of the Hilbert space. The magnetization is transformed into an occupation number $M_A \mapsto \frac{1}{2} - \frac{M_A}{2 n_A}$, with $n_A$ denoting the number of qubits in subsystem $\mathcal{H}_A$.

Fig.~\hyperref[fig:iso_Heisenberg]{\ref*{fig:iso_Heisenberg}(a)} shows results for the two qubit system. The non-locality reaches its maximal value twice during the time interval $t \in \left[ 0, \frac{\pi}{J} \right]$ at which point the time evolution operator oscillates between the swap and the identity operator. In between, $U$ generates entanglement on the trial state which is shown in green. We use QTPD to classically simulate the open quantum dynamics following Eq.~\eqref{eq:open_dynamics}. Starting with the initial state $\ket{1} \otimes \ket{0}$, we simulate the time evolution of the density matrix describing qubit 1 and measure magnetization $M_A$ and entanglement entropy $S$ from Eq.~\eqref{eq:MA} and \eqref{eq:S}. The excitation transfer between the two qubits is reflected in the oscillation of the magnetization between the extremal values 0 and 1. At these points, the entanglement entropy reaches zero indicating an oscillatory swap between $\ket{10}$ and $\ket{01}$. The data from QTPD is in exact agreement with the analytical expressions derived in App.~\ref{sec:Heisenberg}.

On a $3\times 2$ qubit grid on which excitations can swap between neighboring qubits, the overall dynamics is more complicated. While the non-locality of the time evolution operator quickly rises close to the maximum value, it no longer returns to zero within the time interval $t \in [0, \frac{4\pi}{J}]$. The trial initial state, $\ket{110000}$, that is evolved in time, does not return to a product state in the considered time interval, which shows that part of the non-zero non-locality is in fact entanglement generated by $U$. We discuss two measures of entangling power (see App.~\ref{app:entanglement_gen}) on the example of the Heisenberg model in App.~\ref{sec:Heisenberg} in order to specify the relation between non-locality and the entangling properties of $U$.

\section{Conclusion}
In typical problems of quantum information processing, we are given a quantum circuit unitary $U$ as a quantum resource. A quantum tensor product decomposition enables the separation of a small subsystem $\mathcal{H}_A$ from its environment and captures the effective action of $U$ on $\mathcal{H}_A$. This allows one to classically predict entanglement features of $U$ and post-process mereology and short depth compilation algorithms. Furthermore, it enables the study of open quantum dynamics interpreting the two subsystems, $\mathcal{H}_A$ and $\mathcal{H}_B$, as system and environment, respectively. A generalization to decompose arbitrary tensors does not seem straight-forward (see Appendix \ref{app:arbitraryoperators}). Using block-encoding techniques, QTPD could be applied to arbitrary tensors. A hurdle that arises in that case is taming the sample complexity when sampling from the ancillary qubits needed for the block-encoding of matrix product operators \cite{nibbi2023block}. We remain curious about extending QTPD to include quantum states (possibly via block-encoding) where the sum of the QTPD coefficients, $\sum_k s_k$, may be used as a criterion to detect bipartite entanglement (see ``realignment criterion'' in e.g. \cite{chen2002matrix, Horodecki_2009, G_hne_2009}) or verify matrix product operator structures of density operators \cite{Guth_Jarkovsk__2020}.

QTPD paves the way for entanglement investigations at the operator level. A natural next step is to combine QTPD with an iterative search for quasi-classicality emerging in quantum systems. The non-locality that upper bounds entangling power can be used as a cost function to minimize the growth of entanglement with the environment. Also in reverse, QTPD can be used to verify decoherence free structures \cite{Lidar_2003}.

\change{
	Depending on available resources in memory and connectivity, either the preparation of Bell pairs in depth 2, or the sequential approach can be used in a near-term application including QTPD. For medium-term quantum computing, a doubling in qubit number is within the scaling plans of modern quantum computing architectures, which typically strive for exponential growth. We can thus foresee an application in the near future that, for instance, utilizes 120 qubits (or 60 qubits in sequence) to solve the open dynamics of a 15+45 qubit system, a simulation problem that is no longer accessible with classical methods. One such application is the} integration of QTPD into dynamical mean field methods, for instance for the simulation of impurity models. We leave these directions for future research.

\change{
	Alternative process learning methods \cite{caro2022learning, caro2022outofdistribution, huang2022} make use of quantum machine learning-assisted post-processing. While a comparison of the efficiency is not immediate, QTPD provides a stable solution using tomography instead of quantum optimization. 
	
	Similar to process learning approaches, QTPD can be generalized to capture the action of an arbitrary quantum channel $\mathcal{E}$ using its Choi state. Quantum circuits on near-term hardware will inevitably suffer from noise and therefore a QTPD on a noisy circuit is a natural next step. If hardware allows for the \textit{intended} simulation of a specified quantum channel $\mathcal{E}$, QTPD along with its applications can be straightforwardly generalized substituting the unitary $U$ by $\mathcal{E}$.
}

\begin{acknowledgments}
    RM thanks Norbert Schuch, Martin Larocca and Dhrumil Patel for valuable discussions. This work received support from the German Federal Ministry of Education and Research via the funding program, Quantum Technologies - From Basic Research to the Market, under contract number 13N16067 “EQUAHUMO”. It is also part of the Munich Quantum Valley, which is supported by the Bavarian state government with funds from the Hightech Agenda Bayern Plus. ZH acknowledges support from the Sandoz Family Foundation-Monique de Meuron program for Academic Promotion. AA and ATS acknowledge support from the U.S. Department of Energy (DOE), Office of Science, Office of High Energy Physics, QuantISED program. This work was supported by the U.S. Department of Energy (DOE) through a quantum computing program sponsored by the Los Alamos National Laboratory Information Science \& Technology Institute. The LANL approval code for this paper is: LA-UR-24-21093.
\end{acknowledgments}

\bibliographystyle{apsrev4-2}
\bibliography{BiblioR, quantum}

\onecolumngrid
\newpage
\appendix

\section{Ambiguities in the Tensor Decomposition}
\label{app:ambiguities}
The tensor product decomposition is non-unique. Every tensor product basis representation of operators in $\mathcal{H}_{A/B}$ gives a decomposition of the form of Eq.~\eqref{eq:U_schmidt}, not necessarily, but potentially, with minimal rank, $R_\mathrm{min}$. Using the Gram-Schmidt theorem, we can assume the basis $\{A_k \otimes B_k \}_k$ that appears in Eq.~\eqref{eq:U_schmidt} to be orthogonal with respect to the Hilbert-Schmidt inner product, i.e.
\begin{align}
	\braket{A_k \otimes B_k | A_l \otimes B_l} &= \braket{A_k | A_l} \braket{B_k | B_l} \nonumber \\
	&\stackrel{!}{=} \delta_{kl} \norm{A_k}^2 \norm{B_k}^2.
	\label{eq:GS_cond}
\end{align}
Note that this ensures orthogonality of some of the $A_k$ or $B_k$, but not all. Let us assume for now that $\braket{B_k|B_l} = \delta_{kl} \norm{B_k}^2$, so that Eq.~\eqref{eq:GS_cond} is satisfied. A second symmetry is scale invariance, $A_k \otimes B_k = (\lambda A_k) \otimes \left( \frac{1}{\lambda} B_k \right)$, for $\lambda \in \mathds{C}$. With this, we can, for instance, fix the norms of the $B_k$ to be equal to the dimension of $\mathcal{H}_B$, i.e. $\norm{B_k}^2 = d_B$. Moreover, we can rotate the $A_k$ and simultaneously counter-rotate the $B_k$ with a linear superoperator $L: \Span\{A_k\} \to \Span\{A_k\}$ that maps $L(A_k) = \sum_j L_{jk} A_j$. $L$ can also be interpreted as a superoperator on $\Span\{B_k\}$ with the analogue action $L(B_k) = \sum_j L_{jk} B_j$. If we assume $L$ to be unitary, i.e. $\sum_k L_{ik} L_{jk}^* = \sum_k L_{ki} L_{kj}^* = \delta_{ij}$, we can straight-forwardly see that
\begin{align}
	U &= \sum_k A_k \otimes B_k = \sum_k L^\dagger (L(A_k)) \otimes B_k \nonumber \\
	&= \sum_{i,j,k} L_{ji}^* L_{jk} A_i \otimes B_k \nonumber \\
	&=: \sum_j \tilde A_j \otimes \tilde B_j,
\end{align}
is also a tensor decomposition. We defined $\tilde A_j := \sum_i L_{ji}^* A_i$ and $\tilde B_j := \sum_i L_{ji} B_i$. Whenever the boundary of the sum is omitted, we understand the sum to go through $k \in \{1, ... ,R\}$. Unitarity of $L$ implies that the orthogonality of the $B_k$ remains. Further, one can choose $L$ such that the $A_k$ are also orthogonal. To see this, define the linear operator $\mathcal{O}_{kl} = \braket{A_k|A_l}$ and the transformed version $\tilde{ \mathcal{O}}_{kl} = \braket{\tilde A_k|\tilde A_l}$. Inserting the definition of $\tilde A_k$, we get
\begin{align}
	\tilde{ \mathcal{O}}_{kl} = \sum_{i,j} L_{ki} L_{lj}^* \mathcal{O}_{ij} = (L \mathcal{O} L^\dagger)_{kl} \; .
\end{align}
Since $\mathcal{O}$ is hermitian, we can choose $L^\dagger$ to consist of the eigenvectors of $\mathcal{O}$, such that with this choice
\begin{align}
	\tilde{\mathcal{O}}_{kl} = \braket{\tilde A_k|\tilde A_l} = \delta_{kl} \norm{A_k}^2.
\end{align}
With slight abuse of notation, we omit the tilde in the following and store the information about the norms $\norm{A_k}$ separately in scalars $s_k = \frac{\norm{A_k}}{d_A}$, which normalizes the $A_k$ accordingly. Finally, we can also choose the $s_k$ to be non-negative, real numbers as any complex phase can be absorbed into the $B_k$, for instance
\begin{align}
	s_k e^{i \phi_k} A_k \otimes B_k = s_k A_k \otimes (e^{i \phi_k} B_k).
\end{align}
This transformation does not alter the norms and orthogonality of the $B_k$. We have now chosen a specific tensor decomposition $U = \sum_k s_k A_k \otimes B_k$, for which we can assume the following without loss of generality:
\begin{enumerate}
	\item The operators $A_k$ and $B_k$ are orthogonal with respect to the Hilbert-Schmidt inner product
	\item The $A_k$ and $B_k$ are normalized, such that $\norm{A_k}^2 = d_A$ and $\norm{B_k}^2 = d_B$
	\item The $s_k$ are non-negative, real numbers
\end{enumerate}
By unitarity of $U$, the $s_k$ are further constrained to sum up to one, i.e. $\sum_k s_k^2 = 1$.

\section{Two-Step Quantum Tensor Product Decomposition}
In the following, we provide technical details for the proposed algorithm. QTPD involves two steps. First, the operators $A_k$ acting on the smaller subsystem are obtained classically together with the coefficients $s_k$ via diagonalization of a tomographic snapshot of a Choi state. In the second step, this information is used to construct a projective measurement that allows the distillation of $B_k$ out of $U$.

\subsection{Classical snapshot state}
\label{app:snapshot}
The first step of QTPD involves a tomography of the density matrix $\rho_A(U)$ of the Choi-state of $U$ reduced to subsystem $A$, which we calculate in the following. First, define the full Choi state
\begin{equation}
	|\Phi(U)\rangle := (\mathbb{I} \otimes U \otimes \mathbb{I}) \ket{\Phi^+_A} \ket{\Phi^+_B} = \sum_{k}^R \sum_{i_A, i_B, }^{d_{A/B}} s_k |i_A\rangle \otimes A_k |i_A\rangle \otimes   |i_B\rangle \otimes B_k |i_B\rangle
\end{equation}
After tracing out $\mathcal{H}_B$, we get
\begin{align}
	\begin{split}
		\tomo{$\ket{\Phi^+_A}$}{$\ket{\Phi^+_B}$}{$\rho_A(U)$}    
	\end{split}
	\qquad
	\begin{split}
		\rho_A(U) &= \Tr_B\left( |\Phi(U)\rangle \langle \Phi(U)| \right) \\
		&= \frac{1}{d_A d_B} \sum_{k,l}^R \sum_{i_A, i_B, j_A, j_B}^{d_{A/B}} s_k s_l \Tr\left( |i_B\rangle \langle j_B| \right) \Tr\left(  B_k | i_B\rangle \langle j_B | B_l^\dagger \right) |i_A\rangle \langle j_A| \otimes A_k |i_A\rangle \langle j_A|A_l^\dagger   
	\end{split}
	\label{eq:snap}
\end{align}
Executing the trace and using the orthonormality of the $B_k$ yields
\begin{align}
	\rho_A(U) &= \frac{1}{d_A d_B} \sum_{k,l}^R \sum_{i_A, i_B, j_A, j_B}^{d_{A/B}} s_k s_l \, \delta_{i_{B}j_{B}} \, \langle j_B | B_l^\dagger B_k | i_B\rangle \,  |i_A\rangle \langle j_A| \otimes A_k |i_A\rangle \langle j_A|A_l^\dagger
	\nonumber \\
	&= \frac{1}{d_A d_B} \sum_{k,l}^R \sum_{i_A, j_A}^{d_A} s_k s_l \braket{B_l | B_k} \quad |i_A\rangle \langle j_A| \otimes A_k|i_A\rangle \langle j_A| A_l^\dagger \nonumber \\
	&= \frac{1}{d_A} \sum_{k}^R s_k^2 \sum_{i_A, j_A}^{d_A} |i_A\rangle \langle j_A| \otimes A_k|i_A\rangle \langle j_A|  A_k^\dagger 
\end{align}
Finally, we identify $\mathrm{vec}(A_k) = \frac{1}{\sqrt{d_A}} \sum_i^{d_A} \ket{i} \otimes A_k \ket{i}$ which yields the result. Since the Hilbert-Schmidt product and the Euclidean inner product of vectorized operators are the same, we can read off the eigenvectors and eigenvalues of $\rho_A(U)$
\begin{align}
	\rho_A(U) \cdot \mathrm{vec}(A_m) = \frac{1}{d_A} &\sum_k s_k^2 \mathrm{vec}(A_k) \braket{A_k|A_m} = \frac{s_m^2}{d_A} \mathrm{vec}(A_m).
\end{align}

\subsection{B-Distillation}
\label{app:distillation}
To distill the action of $U$ on the larger subsystem $\mathcal{H}_B$, we need to measure the projector $P_k = A_k \ket{\Phi^+_A} \bra{\Phi^+_A} A_k^\dagger$ on $\mathcal{H}_A$. \changeTwo{The projective property of the $P_k$ directly follows from $P_k^\dagger = P_k$ and $P_k^2 = P_k$.} The output state of of the distillation circuit can be straight-forwardly derived in graphical notation
\begin{align}
	\begin{split}
		\distill{$\ket{\Phi^+_A}$}{$\ket{\psi}$}{$\frac{B_k \ket{\psi}}{\norm{B_k \ket{\psi}}}$}
	\end{split}
	\qquad
	\begin{split}
		\frac{1}{d_A} \sum_k^R s_k \vcenter{\hbox{\distillout}} = s_m B_m \ket{\psi}
	\end{split}
	,
	\label{eq:dist_graph}
\end{align}
where we used the orthonormality of the $A_k$. The factor $s_m$ arises as we are not dealing with a unital channel. Measuring $P_m$ involves post-selection on one of the generalized Bell states $\ket{\Phi^+_A}$. We can also use an algebraic formula to prove the above statement. The partial measurement of the $A$-subsystem yields the (unnormalized) state
\begin{align}
	P_m U \ket{\Phi^+_A} \ket{\psi} &= \sum_k^R s_k A_m \ket{\Phi^+_A} \bra{\Phi^+_A} A_m^\dagger A_k \ket{\Phi^+_A} B_k \ket{\psi} = \sum_k^R s_k A_m \ket{\Phi^+_A} \frac{\braket{A_m | A_k}}{d_A} B_k \ket{\psi} \nonumber \\
	&= s_m A_m \ket{\Phi^+_A} B_m \ket{\psi},
	\label{eq:distill_out}
\end{align}
where we used $\norm{A_k}^2 = d_A$. We arrive at the same result as in Eq.~\eqref{eq:dist_graph} modulo tracing out $\mathcal{H}_A$. \changeTwo{In practice, we want to construct a single measurement that outputs the states of Eq.~\eqref{eq:distill_out} for all $m\in \{1, ..., R\}$. This measurement can be described by the quantum channel $\mathcal{E}\left[ \rho \right] = \sum_m^R P_m \rho P_m + Q \rho Q$, where $Q=\mathds{1} - \sum_m P_m$ is necessary to make $\mathcal{E}$ trace preserving, however the action of $Q$ on the input state of Eq.~\eqref{eq:dist_graph}, $\rho = U \left( \ket{\Psi^+_A}\bra{\Phi^+_A} \otimes \ket{\psi} \bra{\psi} \right) U^\dagger$, vanishes. The output state is thus the following mixed state
	\begin{align}
		\mathcal{E}[\rho] = \sum_m s_m^2 \left( A_m \ket{\Psi^+_A}\bra{\Phi^+_A} A_m^\dagger \right) \otimes \left( B_m \ket{\psi} \bra{\psi} B_m^\dagger \right) \qquad \xrightarrow{\Tr_A(.)} \qquad \sum_m s_m^2 B_m \ket{\psi} \bra{\psi} B_m^\dagger,
	\end{align}
	describing a statistical mixture of the states of Eq.~\eqref{eq:distill_out} with probabilities $p_k = \Tr\left( P_k U (\ket{\Phi_A^+}\bra{\Phi_A^+} \otimes \ket{\psi}\bra{\psi}) U^\dagger P_k \right) = s_k^2 \norm{B_k \ket{\psi}}^2$. Note that from unitarity of $U$ we have $\norm{U (\ket{\Phi_A^+} \otimes \ket{\psi})}^2 = \norm{\ket{\Phi_A^+} \otimes \ket{\psi}}^2 = 1$ and therefore 
	\begin{align}
		1 = \norm{U (\ket{\Phi_A^+} \otimes \ket{\psi})}^2 = \sum_{k,l=1}^R s_k s_l \frac{1}{d_A} \braket{A_k | A_l} \braket{\psi | B_k^\dagger B_l | \psi} = \sum_{k=1}^R s_k^2 \norm{B_k \ket{\psi}}^2 = \sum_{k=1}^R p_k. 
	\end{align}
}

\section{Operator approximations using the singular value decomposition}
To show that a collection of the $A_k$ and $B_k$ also yield an optimal low rank approximation, we need to show that they form a global minimum of Eq.~\eqref{eq:low_rank}. \change{The approximation of $U$ by a rank $k$ tensor decomposition is equivalent to finding a rank $k$ approximation of $\mathcal{R}(U)$ from van Loan and Pitsianis' algorithm}, i.e.
\begin{align}
	\norm{\mathcal{R}(U) - \sum_k^r \change{t_k} \mathrm{vec}(C_k)\mathrm{vec}(D_k)^\dagger}.
	\label{eq:low_rank2}
\end{align}
We will show that \change{$t_k = s_k$}, $C_k = A_k$ and $D_k = B_k \;\forall k$ minimizes Eq.~\eqref{eq:low_rank2}. This statement has been proven in \cite{golub2013matrix} using a distance induced by the spectral norm $\norm{.}_\infty$. The generalization to the Frobenius distance is well known, but proofs are often omitted in the literature. We present one here. 

\subsection{Optimal Low Rank Approximation}
\label{app:lowrank}
\begin{theorem}
	\label{prop:rank_approx}
	Let $T \in \mathcal{L}(\mathcal{H})$ with a singular value decomposition $T = \sum_{i=1}^R \sigma_i u_i v_i^\dagger$. Let $T_r$ be the truncation of $T$ to its $r$ largest singular values, i.e. $T_r = \sum_{i=1}^r \sigma_i u_i v_i^\dagger$. Then
	\begin{align}
		\min_{\rank(S) = r} \norm{T - S} = \norm{T - T_r} = \sqrt{ \sum_{j=r+1}^R \sigma_j^2}.
	\end{align}
	\begin{proof}
		The second equality directly follows from the definition of the Frobenius norm. To show the first equality, consider an arbitrary rank $r$ operator, $S$, and calculate the Frobenius distance
		\begin{align}
			\norm{T-S}^2 = \sum_{i=1}^d \sigma_i(T-S)^2 \geq \sum_{i=1}^{d-r} \sigma_i(T-S)^2, \label{eq:sum_ker}
		\end{align}
		where we denoted the $i^{th}$ singular value of matrix $A$ by $\sigma_i(A)$ and dropped the $r$ smallest singular values to estimate a lower bound. Let $T_{r+i-2}$ be the rank $r+i-2$ approximation as defined above. We have
		\begin{align}
			\sigma_{r+i}(T) = \sigma_1(T - T_{r+i-1}) \leq \sigma_1(T - (T-S)_{i-1} - S_r).
		\end{align}
		Here, we denoted the rank $k$ approximation of $T$ by $T_{k}$. The inequality follows from the fact that the rank of $(T-S)_{i-1} + S_r$ is smaller or equal to the rank of $T_{r+i-1}$. Next, consider the inequality
		\begin{align}
			\sigma_1(A+B) \leq \sigma_1(A) + \sigma_1(B),
		\end{align}
		which is a direct consequence of the triangular inequality of the spectral norm. With this, we have
		\begin{align}
			\sigma_1(T - (T-S)_{i-1} - S_r) &\leq \sigma_1(T - S - (T-S)_{i-1}) + \sigma_1(S - S_r) \nonumber \\
			&= \sigma_{i} (T-S) + \sigma_{r+1}(S)
		\end{align}
		Since $S$ is rank $r$, we know that $\sigma_{r+1}(S) = 0$ and we are left with the overall inequality
		\begin{align}
			\sigma_{r+i}(T) \leq \sigma_i(T - S)
		\end{align}
		Returning to Eq.~\eqref{eq:sum_ker}, we can estimate 
		\begin{align}
			\norm{T - S}^2 \geq \sum_{i=1}^{d-r} \sigma_{r+i}(T)^2 = \sum_{i=r+1}^R \sigma_i(T)^2.
		\end{align}
		Taking the square root of both sides shows the statement. 
	\end{proof}
\end{theorem}

\subsection{Nearest Unitary Approximation}
\label{app:nearestUnitary}
In the extreme case where $U$ is close to a rank 1 operator, i.e. $s_1 \approx 1$ and $s_k \approx 0$ for $k \neq 1$, we expect the dominant product operator $A_1 \otimes B_1$ to be close to a unitary. This means that $A_1$ and $B_1$ have to be close to unitaries $U_A$ and $U_B$, individually. In the following, we find the closest unitaries $U_A, U_B$ to $A_1$ and $B_1$. This is a well-known problem that is solved by setting all singular values to 1. The following proposition holds for arbitrary unitary equivalent norms \cite{Fan55}. For pedagogical reasons, let us review the proof for the 2-norm.
\begin{theorem}
	\label{prop:closest_unitary}
	Let $T \in \mathcal{L}(\mathcal{H})$ with a singular value decomposition $T = U \Sigma V^\dagger$, then
	\begin{align}
		\min_{W^\dagger W = \mathds{1}} \norm{T - W} = \norm{T - U V^\dagger}.
	\end{align}
	\begin{proof}
		From unitary invariance of the 2-norm, we can reformulate the minimization problem as follows
		\begin{align}
			\min_{W^\dagger W = \mathds{1}} \norm{U \Sigma V^\dagger - W} = \min_{W^\dagger W = \mathds{1}} \norm{\Sigma - U^\dagger W V} = \min_{W^\dagger W = \mathds{1}} \norm{\Sigma - W},
		\end{align}
		where we redefined the unitary $W$ in the last step using the closedness of the unitary group. The 2-norm can be calculated explicitly as
		\begin{align}
			\norm{\Sigma - W}^2 &= \Tr(\Sigma^2) + \Tr(W^\dagger W) - 2 \Re \left[ \Tr(\Sigma W) \right] \nonumber \\
			&= d + \sum_k^d \left( \sigma_{k}^2 - 2 \sigma_{k} \Re\left[ W_{kk} \right] \right).
		\end{align}
		where we denoted the singular values by $\Sigma_{kl} = \sigma_k \delta_{kl}$. From unitarity of $W$, we know that $\Re\left[ W_{kk} \right] \leq 1$ and thus
		\begin{align}
			\norm{\Sigma - W}^2 \geq d + \sum_k^d \sigma_k^2 - 2 \sigma_k = \sum_k^d (\sigma_k - 1)^2 = \norm{\Sigma - \mathds{1}}^2 \qquad \forall \, W.
		\end{align}
		Therefore, the closest unitary to $T$ is $UV^\dagger$.
	\end{proof}
\end{theorem}
The nearest unitary approximation can be used to find a fast quantum transform $U_{A^{(1)}} \otimes U_{A^{(2)}} \otimes ... \otimes U_{A^{(M)}}$ that approximates the action of $U$. For a Fast Quantum Transform, we need to iterate the unitary approximation for $M$ many tensor product factors. Doing so, we introduce two types of errors, one by a rank one approximation (cf. Proposition \ref{prop:rank_approx}) and one by the nearest unitary approximation of the dominant components $A_1$ (cf. Proposition \ref{prop:closest_unitary}).
\begin{theorem}
	\label{prop:total_approx}
	Let $U = \sum_{j_1, ..., j_M}^{R_1, ..., R_M} s_{j_1...j_M} A_{j_1}^{(1)} \otimes ... \otimes A_{j_M}^{(M)}$ be the multipartite tensor product decomposition of a unitary $U$, with normalized $\norm{A_{j_m}^{(m)}} = \sqrt{d_{A^{(m)}}} \, \forall m \in\{1, ..., M\}$. Further, let $U_m = V_m W_m^\dagger$ be the nearest unitary approximation of $A_1^{(m)} = V_m \Sigma_m W_m^\dagger$ (cf. Proposition \ref{prop:closest_unitary}) and $\frac{1}{ \sqrt{2 d_{A^{(m)}}} } \norm{A_1^{(m)} - U_m} = \epsilon^{(m)} > 0$, as well as $1-s_{1...1} = \epsilon_s > 0$. Then
	\begin{align}
		\frac{1}{ \sqrt{2 d} } \norm{U - U_1 \otimes ... \otimes U_M} \leq \sqrt{ \epsilon_s } + \sqrt{ \frac{1}{2} \epsilon_s^2 + \sum_{m=1}^M \left( \epsilon^{(m)} \right)^2 }
	\end{align}
	\begin{proof}
		We begin with splitting the error using the triangular inequality
		\begin{align}
			\norm{ U - U_1 \otimes ... \otimes U_M } \leq \norm{s_{1...1} A_1^{(1)} \otimes ... \otimes A_1^{(M)} - U_1 \otimes ... \otimes U_M }  + \norm{\sum_{j_1, ..., j_M \neq (1,...,1)}^{R_1, ..., R_M} s_{j_1...j_M} A_{j_1}^{(1)} \otimes ... \otimes A_{j_M}^{(M)} }.
			\label{eq:error_FQT}
		\end{align}
		Let us consider the two terms separately. First, observe that, by construction, 
		\begin{align}
			\braket{A_1^{(m)} | U_m}_{HS} = \Tr\left( A_1^{(m) \dagger} U_m \right) = \Tr\left( W_m \Sigma W_m^\dagger \right) = \Tr\left( \Sigma \right) = \sum_k^{d_{A_1}} \sigma_k \qquad \forall m \in \{1, ..., M\},
		\end{align}
		which is real, i.e. $\braket{A_1^{(m)} | U_m}_{HS} = \braket{U_m | A_1^{(m)}}_{HS}$. Using this, we can relate the Hilbert-Schmidt product above to the 2-norm error $\norm{ P-Q }^2 = \norm{P}^2 + \norm{Q}^2 - 2 \Re\left( \braket{P|Q}_{HS} \right) $ for any two operators $P,Q$. With this we have
		\begin{align}
			&\frac{1}{2 d} \norm{s_{1...1} A_1^{(1)} \otimes ... \otimes A_1^{(M)} - U_1 \otimes ... \otimes U_M }^2 = \frac{1}{2} \left( 1 + s_{1...1}^2 - 2 \frac{s_{1...1}}{d} \prod_m \braket{ A_1^{(m)} | U_m }_{HS} \right) \nonumber \\
			&= \frac{1 + s_{1...1}^2}{2} - s_{1...1} \prod_m^M \left( 1 - \frac{1}{2 d_{A^{(m)}}} \norm{A_1^{(m)} - U_m}^2 \right).
		\end{align}
		The function $\prod_m (1-x_m)$ is convex in the domain $x_m \in [0,1] \, \forall m$, so we can estimate $\prod_m (1-x_m) \geq 1 - \sum_m x_m$ and arrive at
		\begin{align}
			&\frac{1}{2 d} \norm{s_{1...1} A_1^{(1)} \otimes ... \otimes A_1^{(M)} - U_1 \otimes ... \otimes U_M }^2 \leq \frac{1}{2} (1 - s_{1...1})^2 + \sum_m^M \frac{1}{2 d_{A^{(m)}}} \norm{A_1^{(m)} - U_m}^2 \nonumber \\
			&\leq \frac{1}{2} \epsilon_s^2 + \sum_m^M \left( \epsilon^{(m)} \right)^2 ,
		\end{align}
		where we used $1-s_{1...1} = \epsilon_s$. For the second term of Eq.~\eqref{eq:error_FQT}, we use $1-s_{1...1}^2 = 2 \epsilon_s - \epsilon_s^2 \leq 2\epsilon_s$, which also follows from convexity. Using the orthogonality of the $A_{j_m}^{(m)}$ for fixed $m$, the second term reads
		\begin{align}
			\frac{1}{2d} \norm{ \sum_{j_1, ..., j_M \neq (1,...,1)}^{R_1, ..., R_M} s_{j_1...j_M} A_{j_1}^{(1)} \otimes ... \otimes A_{j_M}^{(M)} }^2 = \frac{1}{2} \sum_{j_1, ..., j_M \neq (1,...,1)}^{R_1, ..., R_M} s_{j_1...j_M}^2 = \frac{1 - s_{1...1}^2}{2} \leq \epsilon_s,
		\end{align}
		where the second equality makes use of the normalization of the $s_k$ and the last inequality is convexity again. This concludes the proof.
	\end{proof}
\end{theorem}

\section{Error Propagation for QTPD}
In this section, we follow the error coming from tomography throughout QTPD and its applications and herewith give faithful bounds on the error of predictions given a fixed sample budget for tomography.

\subsection{Error on Tomography and Distillation}
\label{app:errors_QTPD}
The $A_k$ and $s_k$ are captured from the density matrix $\rho_A(U)$ (cf. Eq.~\eqref{eq:snapshot}) via diagonalization
\begin{align}
	\rho_A(U) = \sum_k s_k^2 \, vec(A_k) vec(A_k)^\dagger = V D V^\dagger,
	\label{eq:rho_diag}
\end{align}
finding the eigenbasis $V_{ij} = vec(A_j)_i$ and the eigenvalues $D_{ij} = s_j^2 \delta_{ij}$. From the orthogonality and completeness of the $A_k$, we can show that $V$ is unitary, i.e. $(V^\dagger V)_{ij} = \frac{1}{d_A} \braket{A_i| A_j} = \delta_{ij}$. To get a classical snapshot of Eq.~\eqref{eq:rho_diag}, a tomography is necessary. State tomography suffers from an error that scales inversely with the number $N_S$ of used 
\change{
	samples $\norm{\epsilon^{(T)}} = \mathcal{O}\left( \sqrt{\frac{d_A^2}{N_S}} \right)$ \cite{Apeldoorn23, nielsen2000quantum}. Subsequent refinements require only a sample number of $\mathcal{O}\left( \rank(\rho) \frac{d_A^2}{\epsilon^2} \right)$ \cite{haah2017sample}, or $\mathcal{O}\left( \frac{d_A^2}{\epsilon^2} \right)$ allowing for a small failure probability \cite{o2016efficient}. A recent improvement has found it is necessary to use at least $\Omega\left( \frac{\rank(\rho) d_A^2}{\epsilon} \right)$
} measurements, and it was also conjectured to be sufficient \cite{yuen2022improved}.

Throughout this paper, we consider the 2-norm $\norm{.}$, which gives an average case error if divided by the square root of the Hilbert space dimension. The following discussion can be done straight forwardly for the operator norm, that gives a measure of the worst case error instead, if dimension factors are correctly accounted for. The difference operator $\epsilon^{(T)}$ between the true reduced density matrix $\rho_A(U)$ and the output 
\begin{align}
	\rho_A(U)^{(T)} = \sum_k s_k^{(T) 2} \, vec(A_k^{(T)}) vec(A_k^{(T)})^\dagger = V^{(T)} D^{(T)} V^{(T)\dagger},
	\label{eq:rho(T)_diag}
\end{align}
of the tomography can then be decomposed into deviation of eigenvalues $\epsilon^{(D)}$ and drift of eigenstates $\epsilon^{(V)}$ in the following way
\begin{align}
	\rho_A(U) = \rho_A(U)^{(T)} + \epsilon^{(T)} = V^{(T)} D^{(T)} V^{(T) \dagger} + \epsilon^{(T)} = (V - \epsilon^{(V)}) (D - \epsilon^{(D)}) (V - \epsilon^{(V)})^\dagger + \epsilon^{(T)}.
\end{align}
Solving for $\epsilon^{(T)}$ gives us a relation between those errors. For the sake of a simple presentation, we will estimate the different errors by their maximum $\epsilon = \max \left( \norm{\epsilon^{(D)}}, \norm{\epsilon^{(V)}} \right)$, then
\begin{align}
	\epsilon^{(T)} &= \epsilon^{(V)} D V^\dagger + VD\epsilon^{(V) \dagger} + V \epsilon^{(D)} V^\dagger + \mathcal{O}(\epsilon^2) \nonumber \\
	\norm{\epsilon^{(T)}} &= \norm{\epsilon^{(V)} D V^\dagger + VD\epsilon^{(V) \dagger} + V \epsilon^{(D)} V^\dagger } + \mathcal{O}(\epsilon^2) \leq \norm{\epsilon^{(V)} D} + \norm{VD\epsilon^{(V) \dagger} V} + \norm{ V \epsilon^{(D)} } + \mathcal{O}(\epsilon^2)  \leq 3 \epsilon + \mathcal{O}(\epsilon^2).
	\label{eq:tomo_error_est}
\end{align}
In the second to last step, we used the triangle inequality and in the last step, we used unitary invariance of the 2-norm, as well as submultiplicativity of the 2-norm and $\norm{D} = 1$. At the end of the day, we are interested in the Schmidt values, $s_k$, and the operators $A_k$ and their predictions, $s_k^{(T)}$ and $A_k^{(T)}$, from $\rho_A(U)^{(T)}$. Define
\begin{align}
	\epsilon^{(S)}_k = s_k^2 - {s_k^{(T)}}^2, \qquad \epsilon_k^{(A)} = A_k - A_k^{(T)},
\end{align}
where $\epsilon^{(S)}_k$ are scalars and $\epsilon^{(A)}_k$ are operators. The index $k$ runs through $\{1, ..., R^{(T)}\}$ with the tomographic estimate $R^{(T)}$ of the rank $R$. In the most na\"ive scenario, $R^{(T)}$ will be close or equal to its maximum $d_A^2$, as every error $\epsilon_k^{(S)} \neq 0$. Typically, one needs to define a threshold (dependent on $N_S$) underneath which $s_k^{(T)}$ are considered zero. We can further relate eigenvalue deviation to the error of the $s_k^{(T)}$
\begin{align}
	\norm{\epsilon^{(D)}} &= \norm{D - D^{(T)}} = \sqrt{\sum_k \left( \epsilon^{(S)}_k \right)^2} \leq d_A \epsilon^{(S)},
\end{align}
where we defined $\epsilon^{(S)} = \max_k \left( \epsilon^{(S)}_k \right)$. Similarly, we can relate the eigenstate drift error to the error of the $A_k^{(T)}$. Per construction, the $A_k^{(T)}$ are normalized to $d_A$ and orthogonal to each other, but admit drift angles that are linear in the operators $\epsilon_k^{(A)}$, to be precise $\braket{A^{(T)}_k | A_l} = \delta_{kl} d_A + \braket{A^{(T)}_k | \epsilon_l^{(A)}}$. We collect those drift angles in the matrix $\epsilon^{(A)}_{jk} := \frac{\braket{A_j^{(T)} | \epsilon_k^{(A)}}}{d_A}$. In general, all entries of $\epsilon^{(A)}$ can be non-zero and of the same order of magnitude.
\begin{align}
	\norm{\epsilon^{(A)}_{k}} &= \norm{A_k - A_k^{(T)}} = \sqrt{2 d_A} \sqrt{1 - \frac{1}{d_A} \Re\left[ \braket{ A_k^{(T)} | A_k } \right] } = \sqrt{ - 2 d_A \Re \left[ \epsilon^{(A)}_{kk} \right] } \label{eq:epskk_epsk} \\
	\norm{\epsilon^{(V)}} &= \norm{V - V^{(T)}} = \sqrt{2} d_A \sqrt{1 - \frac{1}{d_A^3} \sum_{k,l} \Re\left[ \braket{ A_k^{(T)} | A_l } \right] } = \sqrt{ - 2 \sum_{k,l} \Re\left[ \epsilon^{(A)}_{kl} \right] }.
\end{align}
Note that the factor $d_A$ makes up for the scaling of the 2-norm in Hilbert space dimension, while the drift angle matrix $\epsilon^{(A)}_{jk}$ does not. The fact that the errors $\norm{\epsilon_k^{(A)}}$ and $\norm{\epsilon_k^{(V)}}$ involve the real part $\Re$ is due to sensitivity of the norm induced distance measure to global phases. Since a global phase difference, e.g. $D \to e^{i \varphi} D$, does not change the outcomes, we might exchange $- \Re \left[\epsilon^{(A)}_{kl}\right]$ by $\abs{\epsilon^{(A)}_{kl}}$, which takes the minimum over $\phi$, without loss of generality. The faulty $A_k^{(T)}$ are further used to filter out the action of the $B_k$. Instead of measuring the projector $P_k$ from Eq.~\eqref{eq:dist}, we have to use $P_k^{(T)} = A_k^{(T)} \ket{\Phi^+_A} \bra{\Phi^+_A} A_k^{(T) \dagger}$ and get the measurement output
\begin{align}
	P_k^{(T)} U \ket{\Phi_A^+} \ket{\psi} &= \sum_l^R s_l \frac{\braket{A_k^{(T)} | A_l}}{d_A} A_k^{(T)} \ket{\Phi^+_A} \otimes B_l \ket{\psi} = \sum_l^R \left( \delta_{kl} + \frac{\braket{A_k^{(T)} | \epsilon_l^{(A)}}}{d_A} \right) s_l A_k^{(T)} \ket{\Phi^+_A} \otimes B_l \ket{\psi} \nonumber \\
	&= s_k A_k^{(T)} \ket{\Phi_A^+} \otimes B_k \ket{\psi} + \sum_l^R \epsilon_{kl}^{(A)} s_l A_k^{(T)} \ket{\Phi^+_A} \otimes B_l \ket{\psi}
\end{align}
In order to normalize this state, we have to multiply it by 
\begin{align}
	N(\epsilon^{(A)}) &:= \frac{1}{s_k \sqrt{d_A}} \left( \sum_{l,m} \lambda_{lm} \left( \delta_{kl} + \frac{s_l}{s_k} \epsilon^{(A)}_{kl} \right) \left( \delta_{km} + \frac{s_m}{s_k} \epsilon^{(A)*}_{km} \right) \right)^{-\frac{1}{2}} \nonumber \\
	&= \frac{1}{s_k \sqrt{d_A}} \left( \frac{1}{\sqrt{\lambda_{kk}}} - \frac{1}{\sqrt{\lambda_{kk}}} \sum_l \frac{s_l}{s_k} \Re \left( \frac{\lambda_{lk}}{\lambda_{kk}} \epsilon^{(A)}_{kl} \right) \right) + \mathcal{O}\left( \epsilon^2 \right),
\end{align}
where we used the coefficients $\lambda_{kl} = s_k s_l \bra{\psi} B_l^\dagger B_k \ket{\psi}$ from Eq.~\eqref{eq:open_dynamics}. The difference between the normalized state vectors then reads
\begin{align}
	&\left( N(\epsilon^{(A)}) P_k^{(T)} - N(0) P_k \right) (\mathds{1}_A \otimes U) (\ket{\Phi^+_A} \otimes \ket{\psi}) \nonumber \\
	&= - \frac{\epsilon^{(A)}_k \ket{\Phi_A^+}}{\sqrt{d_A}} \otimes \frac{B_k \ket{\psi}}{\sqrt{\lambda_{kk}}} + \frac{A^{(T)}_k \ket{\Phi_A^+}}{\sqrt{d_A}} \otimes \sum_l \sqrt{\frac{\lambda_{ll}}{\lambda_{kk}}} \left( \epsilon^{(A)}_{kl} \frac{s_l}{s_k} \frac{B_l \ket{\psi}}{\sqrt{\lambda_{ll}}} - \Re\left( \epsilon^{(A)}_{kl} \frac{s_l}{s_k} \frac{\lambda_{kl}}{\sqrt{\lambda_{kk} \lambda_{ll}}} \right) \frac{B_k \ket{\psi}}{\sqrt{\lambda_{kk}}} \right) + \mathcal{O}(\epsilon^2)
\end{align}
To get an error measure for the output state of Eq.~\eqref{eq:dist_graph}, we calculate the distance between the normalized states
\begin{align}
	&\norm{ \left( N(\epsilon^{(A)}) P_k^{(T)} - N(0) P_k \right) (\mathds{1}_A \otimes U) (\ket{\Phi^+_A} \otimes \ket{\psi}) } \nonumber \\
	&\leq \frac{\norm{\epsilon^{(A)}_k}}{\sqrt{d_A}} + \norm{\sum_l \sqrt{\frac{\lambda_{ll}}{\lambda_{kk}}} \left( \epsilon^{(A)}_{kl} \frac{s_l}{s_k} \frac{B_l \ket{\psi}}{\sqrt{\lambda_{ll}}} - \Re\left( \epsilon^{(A)}_{kl} \frac{s_l}{s_k} \frac{\lambda_{kl}}{\sqrt{\lambda_{kk} \lambda_{ll}}} \right) \frac{B_k \ket{\psi}}{\sqrt{\lambda_{kk}}} \right)} + \mathcal{O}(\epsilon^2).
	\label{eq:B_dist_distance}
\end{align}
If we assume, for the sake of simplicity, that the matrix of drift angles is diagonal, i.e. $\epsilon^{(A)}_{jk} = \epsilon^{(A)}_{kk} \delta_{jk}$, then the error bound gets
\begin{align}
	&\norm{ \left( N(\epsilon^{(A)}) P_k^{(T)} - N(0) P_k \right) (\mathds{1}_A \otimes U) (\ket{\Phi^+_A} \otimes \ket{\psi}) } \leq \frac{\norm{\epsilon^{(A)}_k}}{\sqrt{d_A}} + \norm{ \Im\left( \epsilon^{(A)}_{kk} \right) \frac{B_k \ket{\psi}}{\sqrt{\lambda_{kk}}} } + \mathcal{O}(\epsilon^2) \nonumber \\
	&\leq \frac{\norm{\epsilon^{(A)}_k}}{\sqrt{d_A}} + \abs{ \epsilon^{(A)}_{kk} } + \mathcal{O}(\epsilon^2) = \left( 1+\frac{1}{\sqrt{2}}\right) \frac{1}{\sqrt{d_A}} \norm{\epsilon_k^{(A)}} + \mathcal{O}(\epsilon^2),
\end{align}
where we used Eq.~\eqref{eq:epskk_epsk}. The same result can be achieved by neglecting the terms $\frac{s_l}{s_k}$ for $l \neq k$, which is valid as long as $s_k$ is a dominant singular value. In summary, the error $\epsilon^{(T)}$ from tomography propagates linearly (in leading order) through the digonalization into $s_k$ and $A_k$, as well as through the distillation of the $B_k$ and can be suppressed with raising the number of shots $N_S$.

\subsection{Worst Case Error}
Although the 2-norm is a natural choice to measure distances on the space of vectorized operators as it is induced from the Hilbert-Schmidt inner product. As we normalized the 2-norm before by a factor of dimension, it represents typical errors. In the following, we leave a short note on the worst case error which is measured by the operator norm instead. Analog to Eq.~\eqref{eq:tomo_error_est}, we can relate the operator norm error of tomography to eigenvalue and drift errors 
\begin{align}
	\norm{\epsilon^{(T)}}_\infty \leq 3 \epsilon_\infty + \mathcal{O}(\epsilon^2),
\end{align}
with $\epsilon_\infty = \max\left\{ \norm{\epsilon^{(V)}}_\infty, \norm{\epsilon^{(D)}}_\infty \right\}$, since the operator norm is also unitary invariant and $\norm{D}_\infty \leq 1$. We can further relate
\begin{align}
	\norm{\epsilon^{(D)}}_\infty &= \max_k \abs{ \epsilon^{(S)}_k } \\
	\norm{\epsilon^{(V)}}_\infty &= \norm{ \epsilon^{(A)} }_\infty,
\end{align}
where we defined the matrix $\epsilon^{(A)}$ with elements $\epsilon^{(A)}_{jk}$. The second identity follows from unitary invariance. Using the calculation from Eq.~\eqref{eq:B_dist_distance}, we can straight-forwardly bound the error on the distillation of the $B_k$ via
\begin{align}
	&\norm{\left( N(\epsilon^{(A)}) P_k^{(T)} - N(0) P_k \right) U \ket{\Phi^+_A} \ket{\psi} } \nonumber \\
	&\leq \frac{\norm{\epsilon_k^{(A)}}}{\sqrt{d_A}} + R \, \max_l \left( \sqrt{\frac{\lambda_{ll}}{\lambda_{kk}}} \frac{s_l}{s_k} \right) \norm{ \epsilon^{(A)}_{kl} }_\infty + \mathcal{O}(\epsilon^2)
\end{align}
which, is loose by a factor of $R \leq d_A^2$, in general, but is reduced in cases where drift is approximately diagonal or only few $s_k$ are dominant. Also here, the tomography error propagates linearly through the errors for $s_k, A_k$ and $B_k$ and can be suppressed with the number of shots $N_S$.

\subsection{Error on Applications}
\label{app:errors_applications}
As for the applications of QTPD, how the error propagates depends on the the objective of interest. Let us start with the non-locality $S_A(U) = - \sum_k s_k^2 \log s_k^2$, which only depends on the Schmidt values, $s_k$, and thus the error depends only on $\epsilon^{(S)} = \frac{\norm{\epsilon^{(D)}}}{d_A}$,
\begin{align}
	\abs{S^{(T)}_A(U) - S_A(U)} &= \abs{\sum_k \left( 2 \left( s^{(T)}_k \right)^2 \log s^{(T)}_k - 2 s_k^2 \log s_k \right) } = \abs{\sum_k \log \frac{s_k^{2 s_k^2}}{\left( s^{(T)}_k \right)^{2\left(s^{(T)}_k\right)^2}}} \nonumber \\
	&= \abs{\sum_k \log \frac{\left( \left( s^{(T)}_k \right)^2 + \epsilon^{(S)}_k \right)^{\left( s^{(T)}_k \right)^2 + \epsilon^{(S)}_k}}{\left( s^{(T)}_k \right)^{2\left( s^{(T)}_k \right)^2}}} \nonumber \\
	&= \sum_k \sgn(\epsilon^{(S)}_k) \left( 1 + \left( s^{(T)}_k \right)^2 \log\left( s^{(T)}_k \right)^2 \right) \epsilon^{(S)} + \mathcal{O}\left( \left( \epsilon^{(S)} \right)^2 \right).
\end{align}
In the last step, we linearized the logarithm in $\epsilon_k^{(S)}$. Since the entangling power from Eq.~\eqref{eq:ent_pow_swap} is just a sum of non-locality measures, the error propagates similarly. If we use QTPD for open quantum dynamics, the tomography error also propagates into expectation values of mixed states, $\sigma_A$. For errors of expectation values, it is sufficient to consider the trace distance of the reduced density matrices $T\left( \sigma_A, \sigma_A^{(T)} \right)$, since it upper bounds errors in expectation values of observables \cite{wilde2013quantum}.
\begin{lemma}
	Let $\rho$ and $\sigma$ be density matrices. The difference of the expectation value of an observable $O$ can be bounded in the following way
	\begin{align}
		\abs{ \Tr(O \rho) - \Tr(O \sigma) } \leq \norm{O}_\infty \norm{\rho - \sigma}_1.
	\end{align}
\end{lemma}
We are thus left with the 1-norm induced distance of the reduced density matrix $\sigma_A^{(T)} = \sum_{k,l} \lambda_{kl} A^{(T)}_k \ket{\psi_A} \bra{\psi_A} A^{(T) \dagger}_l$ from $\sigma_A$ (cf. Eq.~\eqref{eq:open_dynamics}). The difference operator reads 
\begin{align}
	\sigma_A - \sigma_A^{(T)} &= \sum_{k,l} \lambda_{kl} \left[ \left( A^{(T)}_k + \epsilon_k^{(A)} \right) \ket{\psi_A} \bra{\psi_A} \left( A^{(T) \dagger}_l + \epsilon_l^{(T) \dagger} \right) - A^{(T)}_k \ket{\psi_A} \bra{\psi_A}  A^{(T) \dagger}_l \right] \nonumber \\
	&= \sum_{k,l} \lambda_{kl} \left[ \epsilon_k^{(A)} \ket{\psi_A} \bra{\psi_A} A^{(T) \dagger}_l + A^{(T)}_k \ket{\psi_A} \bra{\psi_A} \epsilon_l^{(A) \dagger} \right]+ \mathcal{O}\left( \left( \epsilon^{(A)} \right)^2 \right).
\end{align}
We ignored possible errors in the determination of the coefficients $\lambda_{kl}$. Finally, we can estimate an upper bound for the 1-norm error in the following way
\begin{align}
	\norm{\sigma_A - \sigma_A^{(T)}}_1 &\leq \sum_{k,l} \abs{\lambda_{k,l}} \norm{\epsilon_k^{(A)} \ket{\psi_A} \bra{\psi_A} A^{(T) \dagger}_l + A^{(T)}_k \ket{\psi_A} \bra{\psi_A} \epsilon_l^{(A) \dagger}}_1 + \mathcal{O}\left( \left( \epsilon^{(A)} \right)^2 \right) \nonumber \\ 
	&\leq 2 \sum_{k,l} \abs{\lambda_{k,l}} \norm{\epsilon_k^{(A)} \ket{\psi_A} \bra{\psi_A} A^{(T) \dagger}_l }_1 + \mathcal{O}\left( \left( \epsilon^{(A)} \right)^2 \right) \nonumber \\
	&\leq 2 \sum_{k,l} \abs{\lambda_{k,l}} \abs{ \bra{\psi_A} A^{(T)}_l A^{(T) \dagger}_l \ket{\psi_A} } \norm{\epsilon_k^{(A)} }_2 + \mathcal{O}\left( \left( \epsilon^{(A)} \right)^2 \right) \nonumber \\
	&\leq \left( 2 \sum_{k,l} \abs{\lambda_{k,l}} \abs{ \bra{\psi_A} A^{(T)}_l A^{(T) \dagger}_l \ket{\psi_A} } \right) \max_k \epsilon^{(A)}_k + \mathcal{O}\left( \left( \epsilon^{(A)} \right)^2 \right),
\end{align}
where we used the triangle inequality in the first and second step together with $\norm{A}_1 = \norm{A^\dagger}_1$. In the third step, the Hölder inequality, $\norm{AB}_1 \leq \norm{A}_2 \norm{B}_2$, was used.

\section{Entanglement Generation from QTPD}
\label{app:entanglement_gen}
One measure that singles out the non-local, but also non-entangling action of the swap operator has been introduced in \cite{Jonnadula_2020} for the case of equally large subsystems, i.e. $d_A = d_B$
\begin{align}
	e_A(U) = \frac{1}{\log(d_A^2)} \left( S_A(U) + \left( S_A(U P_{AB}) - \log\left( d_A^2 \right) \right) \right),
	\label{eq:ent_pow_swap_equal}
\end{align}
where $P_{AB}$ swaps the two subsystems, i.e. $P_{AB} \ket{\psi}_A \ket{\psi}_B = \ket{\psi}_B \ket{\psi}_A$. Since we are considering the asymmetrical case $d_A \leq d_B$, let us define a straight-forward generalization in which we sum over all different contributions from permutations between $A$ and subsystems of dimension $d_A$ in $B$
\begin{align}
	e_A(U) = \frac{1}{\log(d_A^2)} \left( S_A(U) + \sum_{\substack{ C \subset B \\ \dim(C) = d_A } } \left( S_A(U P_{AC}) - \log\left( d_A^2 \right) \right) \right),
	\label{eq:ent_pow_swap}
\end{align}
where the sum over subsystems $C$ only ranges over qubit configurations, and is therefore finite. An alternative measure for entangling power is the mean entanglement that is generated by the action of $U$ on product states
\begin{align}
	e_m(U) = \mathds{E}_{\ket{\psi_A}, \ket{\psi_B}} \left[ E(\Tr_B \left( U \ket{\psi_A} \otimes \ket{\psi_B}) \right) \right],
	\label{eq:ent_pow}
\end{align}
where $E$ is an (a priori unspecified) measure of entanglement and $\mathds{E}$ denotes the Haar measure over the subsystems $\mathcal{H}_A$ and $\mathcal{H}_B$. For the linearized entanglement entropy $E(\rho) = 1 - \Tr(\rho^2)$, the entangling power from Eq.~\eqref{eq:ent_pow} has been discussed by Zanardi et al. \cite{Zanardi_2000}. The mean linear entanglement entropy growth from the action of a unitary $U$ reads (cf. Eq.~(5) of \cite{Zanardi_2000}) 
\begin{align}
	e_m(U) = 1 - \frac{1}{d_A (d_A + 1)} \frac{1}{d_B (d_B + 1)} \left( d_A^2 d_B + d_A d_B^2 + \Tr\left( (U \otimes U) P_A (U^\dagger \otimes U^\dagger) P_A \right) + \Tr\left( (U \otimes U) P_B (U^\dagger \otimes U^\dagger) P_A \right) \right),
\end{align}
using two copies of the full system $\mathcal{H}_A \otimes \mathcal{H}_B \otimes \mathcal{H}_A \otimes \mathcal{H}_B$, where $P_{A} \ket{\psi_A} \otimes \ket{\psi_B} \otimes \ket{\phi_A} \otimes \ket{\phi_B} = \ket{\phi_A} \otimes \ket{\psi_B} \otimes \ket{\psi_A} \otimes \ket{\phi_B}$ swaps the states in the two copies of $\mathcal{H}_A$ and analogously does $P_B$ on $\mathcal{H}_B$. Note the difference to the swap operators $P_{AC}$ that were used previously. We can simplify the trace terms by inserting $U = \sum_k s_k A_k \otimes B_k$. For the first term, we get in tensor network notation
\begin{align}
	\vcenter{\hbox{\EmuTermOne}} &= \sum_{k,l,m,n} s_k s_l s_m s_n \vcenter{\hbox{\EmuTermOneAB{A}{l}{n}}} \times \vcenter{\hbox{\EmuTermOneAB{B}{n}{l}}} \nonumber \\
	&=\sum_{k,l,m,n} s_k s_l s_m s_n \delta_{mn} \delta_{kl} \delta_{kn} \delta_{ml} d_A^2 d_B^2 = \sum_k s_k^4 d_A^2 d_B^2,
\end{align}
where the second step makes use of the orthonormality of the $A_k$ and $B_k$. The second trace can be calculated analogously 
\begin{align}
	\vcenter{\hbox{\EmuTermTwo}} &= \sum_{k,l,m,n} s_k s_l s_m s_n \vcenter{\hbox{\EmuTermTwoA{A}{l}{n}}} \times \vcenter{\hbox{\EmuTermTwoB{B}{l}{n}}} \nonumber \\
	&= \sum_{k,l,m,n} s_k s_l s_m s_n \Tr\left( A_k A_l^\dagger A_m A_n^\dagger \right) \Tr\left( B_k B_n^\dagger B_m B_l^\dagger \right).
\end{align}
In summary, the mean entanglement generation reads
\begin{align}
	e_m(U) = 1 &- \frac{d_A + d_B}{(d_A + 1)(d_B + 1)} - \frac{d_A d_B \sum_k s_k^4}{(d_A + 1) (d_B + 1)} \nonumber \\
	&- \frac{1}{d_A (d_A + 1)} \frac{1}{d_B (d_B + 1)} \sum_{k,l,m,n} s_k s_l s_m s_n \Tr\left( A_k A_l^\dagger A_m A_n^\dagger \right) \Tr\left( B_k B_n^\dagger B_m B_l^\dagger \right).
	\label{eq:emu}
\end{align}
The last term cannot be calculated without tomographic knowledge of the $B_k$ and is hence out of reach for near-term quantum computing. It could be solved by a fault-tolerant device. We leave this for future work.

\section{Decoherence Free Structures}
To support the discussion on using QTPD for mereology, we give an example for a product operator transformed into an entangled basis and furthermore give a characterization of unitaries that admit such a basis in which the action is non-entangling.

\subsection{Example for Growth of Operator Entanglement}
\label{app:entanglementgrowth}
Consider a two-qubit Hilbert space $\mathcal{H} \cong \mathcal{H}_A \otimes \mathcal{H}_B$ with a non-entangling T-gate $U= \mathds{1} \otimes T$, $T = \begin{bmatrix}
	1 & 0 \\
	0 & e^{i \frac{\pi}{4}}
\end{bmatrix}$. Further consider the unitarily equivalent gate $VUV^\dagger$ that is related to $U$ by a CNOT rotation, $V = \ket{0}\bra{0} \otimes \mathds{1} + \ket{1}\bra{1} \otimes X = V^\dagger$. One can derive
\begin{align}
	VUV^\dagger = \ket{0}\bra{0} \otimes T + e^{i \frac{\pi}{4}} \ket{1} \bra{1} \otimes T^\dagger.
\end{align}
To show that this is an entangling gate, we can calculate the overlaps with Pauli operators in subsystem $\mathcal{H}_A$:
\begin{align}
	\Tr(VUV^\dagger) &= T + e^{i \frac{\pi}{4}} T^\dagger = (1+e^{i \frac{\pi}{4}}) \mathds{1} \\
	\Tr(Z_A VUV^\dagger) &= T - e^{i \frac{\pi}{4}} T^\dagger = (1-e^{i \frac{\pi}{4}}) Z \\
	\Tr(X_A VUV^\dagger) &= 0 \\
	\Tr(Y_A VUV^\dagger) &= 0 
\end{align}
Altogether, we deduce $VUV^\dagger = (1+e^{i \frac{\pi}{4}}) \mathds{1} + (1-e^{i \frac{\pi}{4}}) Z \otimes Z$.

\subsection{A Necessary and Sufficient Condition for the existence of Decoherence Free Subsystems}
\label{app:decfreesub}
\begin{theorem}
	Let $U \in \mathcal{L}(\mathcal{H})$ be a unitary with eigenstates $\ket{\theta_i}$, i.e. $U \ket{\theta_i} = e^{i \theta_i} \ket{\theta_i}$. Further, let $\mathcal{H} \cong \mathcal{H}_A \otimes \mathcal{H}_B$ with $\dim(\mathcal{H}_A) =:d_A < d_B := \dim(\mathcal{H}_B)$ define a split indexed by $i = (\mu, m), \mu \in \{1, ..., d_A\}, m \in \{1, ..., d_B\}$ and $i \in \{1,..., d_A d_B\}$. Then
	\begin{align}
		\exists \quad \textrm{ ordering} \quad \theta_i = \theta_{\mu m} &= \phi_\mu + \psi_m \quad \textrm{ for some } \quad \phi_\mu, \psi_m \in [0,2\pi) \nonumber \\
		&\iff \nonumber \\
		\exists V \quad \textrm{unitary, s.t.} \quad VUV^\dagger &= A \otimes B \quad \textrm{with} \quad A, B \quad \textrm{unitary}
	\end{align}
	\begin{proof}
		The backwards direction ``$\Leftarrow$" becomes trivial as soon as we write down $U$ in diagonal form. Let $T$ denote the eigenbasis of $U$, i.e.
		\begin{align}
			U = T^\dagger D_U T = T^\dagger \begin{bmatrix}
				e^{i \theta_1} && \\
				& \ddots &\\
				& & e^{i \theta_d}
			\end{bmatrix}
			T = V^\dagger (A \otimes B) V \quad \iff \quad A \otimes B = VT^\dagger D_U T V^\dagger.
		\end{align}
		Thus, $U$ and $A\otimes B$ share the same eigenvalues. If we denote the eigenvectors of $A$ by $\ket{\phi_\mu}$ and of $B$ by $\ket{\psi_m}$, we get $(A \otimes B) \ket{\phi_\mu} \otimes \ket{\psi_m} = e^{i (\phi_\mu + \psi_m)} \ket{\phi_\mu} \otimes \ket{\psi_m}$ and thus $\theta_{\mu m} = \phi_\mu + \psi_m$.
		
		For the forward direction ``$\Rightarrow$" consider again the eigenbasis $V$ of $U$. As we know that the eigenvalues are related by $\theta_i = \phi_\mu + \psi_m$, we can put $V$ into an order such that $D_U = D_A \otimes D_B$ decouples with
		\begin{align}
			D_A = \begin{bmatrix}
				e^{i \phi_1} && \\
				& \ddots &\\
				& & e^{i \phi_{d_A}}
			\end{bmatrix} \quad \textrm{and} \quad D_B = \begin{bmatrix}
				e^{i \psi_1} && \\
				& \ddots &\\
				& & e^{i \psi_{d_B}}
			\end{bmatrix}.
		\end{align}
		We conclude with $U = V^\dagger (D_A \otimes D_B) V$.
	\end{proof}
\end{theorem}

\section{Generalization to arbitrary operators}\label{app:arbitraryoperators}
One might be interested in a tensor decomposition of a non-unitary operator $T \in \mathcal{L}(\mathcal{H}_A \otimes \mathcal{H}_B)$, for instance a Hermitian operator. QTPD can be generalized to general non-unitary operators by utilizing the concept of block encodings \cite{Gily_n_2019}. If $U$ is an $(\alpha, a, \epsilon)$ block-encoding of $T$, we can use the same circuits as in Eq.~\eqref{eq:snap} and Eq.~\eqref{eq:dist_graph} together with post-selection on $\ket{0...0}$ on the ancillary system, i.e.
\begin{align}
	\tomoblock{$\ket{\Phi^+_A}$}{$\ket{\Phi^+_B}$}{$\rho_{T|A}$}{$\ket{0}$}{$\ket{0}$} \qquad \qquad \distillblock{$\ket{\Phi^+_A}$}{$\ket{\psi}$}{$\frac{B_m \ket{\psi}}{\norm{B_m \ket{\psi}}}$}{$\ket{0}$}{$\ket{0}$}
	\label{eq:non-unitary-circuits}
\end{align}
The generalization to non-unitary operators $T$ comes at a price of raising the sample complexity in the two circuits of Eq.~\eqref{eq:non-unitary-circuits}. In particular, the sample number will be multiplied by a factor exponential in the number of ancilla qubits $2^a$. This puts a restriction onto the tensors $T$ that can be analyzed this way. While there are typically upper bounds for $a$ polynomial in the qubit number $n$ \cite{Gily_n_2019}, one would need a block encoding with $a = \mathcal{O}(1)$ in order to keep the sample complexity below full tomography.

\section{Analytical Discussion of the Heisenberg Model on Two Qubits}
\label{sec:Heisenberg}
Consider the Hamiltonian of the Heisenberg model for two qubits and the time evolution operator $U$ that we consider as a black box unitary for tensor decomposition
\begin{align}
	H = - \left( J_x \, X_1 X_2 + J_y \, Y_1 Y_2 + J_z \, Z_1 Z_2 \right) \qquad \text{and} \qquad U := e^{-itH} = e^{i J_x t X_1 X_2}e^{i J_y t Y_1 Y_2}e^{i J_z t Z_1 Z_2}. \label{eq:Heisenberg_2q} 
\end{align}
In the following, we perform all calculations with distinct interaction strengths $J_x, J_y$ and $J_z$ and view the isotropic case as an example where $J_x = J_y = J_z =: J$. The time evolution operator $U$ decays into local exponentials, because -- on 2 qubits -- all terms in the Hamiltonian commute. With the identity for Pauli exponentials $e^{iJ_x t X_1 X_2} = \cos(J_x t) \mathds{1} + i \sin(J_x t) X_1 X_2$ (and similar for the other two Pauli-strings), we can directly calculate the tensor decomposition of $U$
\begin{align}
	U &= \left( c_x \mathds{1} + i s_x X_1 X_2 \right) \left( c_y \mathds{1} + i s_y Y_1 Y_2 \right) \left( c_z \mathds{1} + i s_z Z_1 Z_2 \right) \nonumber \\
	&= \left( c_x c_y c_z + i s_x s_y s_z \right) \mathds{1} + \left( i s_x c_y c_z + c_x s_y s_z \right) X_1 X_2 + \left( i s_y c_x c_z + c_y s_x s_z \right) Y_1 Y_2 + \left( i s_z c_y c_x + c_z s_y s_x \right) Z_1 Z_2 \nonumber \\
	&=: g_0 \mathds{1} + g_x X_1 X_2 + g_y Y_1 Y_2 + g_z Z_1 Z_2
	\label{eq:U_dec}
\end{align}
We introduced the shorthand notation $c_i := \cos(J_i t)$ and $s_i = \sin(J_i t) \quad \forall i \in \{x,y,z\}$ and omitted the time-dependence in the following for the sake of clarity. We also allow ourselves some flexibility in pushing complex phases between $s_k$ and $B_k$, which is technically against our convention introduced in App.~\ref{app:ambiguities} assuming the $s_k$ to be real. With Eq.~\eqref{eq:U_dec}, we can read off the Schmidt coefficients $g_i$ and deduce that $U$ has maximal Schmidt rank 4 except for when one of the terms vanishes. In the isotropic case, this happens for $t = \frac{\pi}{2 J}$. If we require $J_z = 0$ and fix $J_x = J_y = J$, we recover the swap gate $S$ at $t = \frac{\pi}{4 J}$, i.e.
\begin{align}
	S = e^{ i \frac{\pi}{4} (X_1 X_2 + Y_1 Y_2)} = \frac{1}{2} \left( \mathds{1} + X_1 X_2 + Y_1 Y_2 + Z_1 Z_2 \right).
\end{align}
Hence, we get the tensor decomposition of the swap gate for free from Eq.~\eqref{eq:U_dec}. We can use this to write down the entangling power of $U$ on qubit 1
\begin{align}
	e_1(U) &= \frac{1}{\log(4)} \left( S_1(U) + S_1(US) \right) - 1. \label{eq:e1U} \\
	S_1(U) &= - \sum_k \abs{g_k}^2 \log \abs{g_k}^2 \\
	S_1(US) &= - \frac{1}{4} \abs{g_0 + g_x + g_y + g_z}^2 \log\frac{\abs{g_0 + g_x + g_y + g_z}^2}{4} - \frac{1}{4}\abs{g_0 + g_x - g_y - g_z}^2 \log\frac{\abs{g_0 + g_x - g_y - g_z}^2}{4} \nonumber \\
	&- \frac{1}{4}\abs{g_0 - g_x + g_y - g_z}^2 \log\frac{\abs{g_0 - g_x + g_y - g_z}^2}{4} - \frac{1}{4}\abs{g_0 - g_x - g_y + g_z}^2 \log\frac{\abs{g_0 - g_x - g_y + g_z}^2}{4}.
\end{align}
One can see directly that for $U=S$, the entangling power vanishes while the non-locality measure $S_1(U)$ is maximal. In order to get large entangling power, both $S_1(U)$ and $S_1(US)$ have to be large. The mean entanglement generation, that is derived in App.~\ref{app:entanglement_gen}, behaves similarly in the 2-qubit case. The last term in Eq.~\eqref{eq:emu} is the only non-trivial term to evaluate. The only non-vanishing trace terms yield
\begin{align}
	\sum_{k,l,m,n} g_k g_l^* g_m g_n^* \Tr\left( A_k A_l^\dagger A_m A_n^\dagger \right) \Tr\left( B_k B_n^\dagger B_m B_l^\dagger \right) = &d^2 \sum_k \abs{g_k}^4 \nonumber \\
	+ &d^2 \sum_{k \neq l} \left( 2\abs{g_k}^2 \abs{g_l}^2 + g_k^2 g_l^{2*} \right) \nonumber \\
	+ &d^2 \sum_{\sigma \in S_4} g_{\sigma(0)} g_{\sigma(x)}^* g_{\sigma(y)} g_{\sigma(z)}^*,
\end{align}
with $d=2$. The three terms arise from different combinations of inserting the Pauli operators $A_k, B_k \in \{\mathds{1}, X, Y, Z\}$. Since Pauli operators are traceless, the products $A_k A_l^\dagger A_m A_n^\dagger$ have to result into $\mathds{1}$, such that the trace yields a factor of dimension $d$. The first term comes from traces of the form $\Tr(A_k^4)^2$, the second from $\Tr(A_k^2 A_l^2)^2$, as well as $\Tr(A_k A_l A_k A_l)^2$ with $k \neq l$ and adequate combinatorial coefficients. Finally the third term represents traces in which all operators are different, i.e. $\Tr(A_k A_l A_m A_n) \Tr(A_k A_n A_m A_l) = - \Tr(A_k A_l A_m A_n)^2$ with $(k,l,m,n) = (\sigma(0), \sigma(x), \sigma(y), \sigma(z)), \sigma \in S_4$.

Now let us fix the initial state to be $\ket{10}$. The effective open time evolution of the first qubit under the Hamiltonian $H$ (Eq.~\eqref{eq:Heisenberg_2q}) can be expressed as a quantum channel $\mathcal{E}_t$ that evolves a density matrix describing the quantum state of qubit 1
\begin{align}
	\mathcal{E}_t(\ket{1}\bra{1}) = \rho_1 (t) &= (\abs{g_0}^2 + \abs{g_z}^2 - g_0 g_z^* - g_z g_0^* ) \ket{1}\bra{1} + (\abs{g_x}^2 + \abs{g_y}^2 + g_x g_y^* + g_y g_x^* ) \ket{0}\bra{0} \nonumber \\
	&= \abs{g_0 - g_z}^2 \ket{1} \bra{1} + \abs{g_x + g_y}^2 \ket{0} \bra{0} \nonumber \\
	&= \cos^2((J_x + J_y)t) \ket{1}\bra{1} + \sin^2((J_x + J_y)t) \ket{0}\bra{0}.
	\label{eq:10_evol}
\end{align}
We can read off the Schmidt values of this state $\abs{\lambda_0}^2 = \sin^2((J_x + J_y)t)$ and $\abs{\lambda_1}^2 = \cos^2((J_x + J_y)t)$. In order to learn properties of the output state, we measure the magnetization $\braket{Z}_{\rho_1(t)}$ and the entanglement entropy $S(\rho_1(t))$ of the time-evolved state in the numerical experiment. In the two-qubit case, we can write down the two observables as functions of time
\begin{align}
	\braket{Z}_{\rho_1(t)} &= \sin^2((J_x + J_y)t) - \cos^2((J_x + J_y)t) \\
	S(\rho_1(t)) &= \sin^2((J_x + J_y)t) \log\left( \sin^2((J_x + J_y)t) \right) + \cos^2((J_x + J_y)t) \log\left( \cos^2((J_x + J_y)t) \right).
	\label{eq:Sr1t}
\end{align}
The above example starts with a product state $\ket{10}$ and does not show any quantum coherence after time evolution. If we start with a product state $\ket{1+}$, which lies skew in two spin symmetry sectors, some of the coherence on qubit 2 gets swapped to qubit 1
\begin{align}
	\mathcal{E}_t(\ket{1}\bra{1}) = \rho_1 (t) &= (\abs{g_0}^2 + \abs{g_z}^2 ) \ket{1}\bra{1} + (\abs{g_x}^2 + \abs{g_y}^2) \ket{0}\bra{0} + \left( g_z g_y^* - g_0 g_x^*  \right) \ket{1}\bra{0} +\left(  g_y g_z^* - g_x g_0^* \right) \ket{0}\bra{1},
\end{align}
as opposed to Eq.~\eqref{eq:10_evol}. 
\begin{figure*}
	\centering
	\includegraphics[scale=0.35]{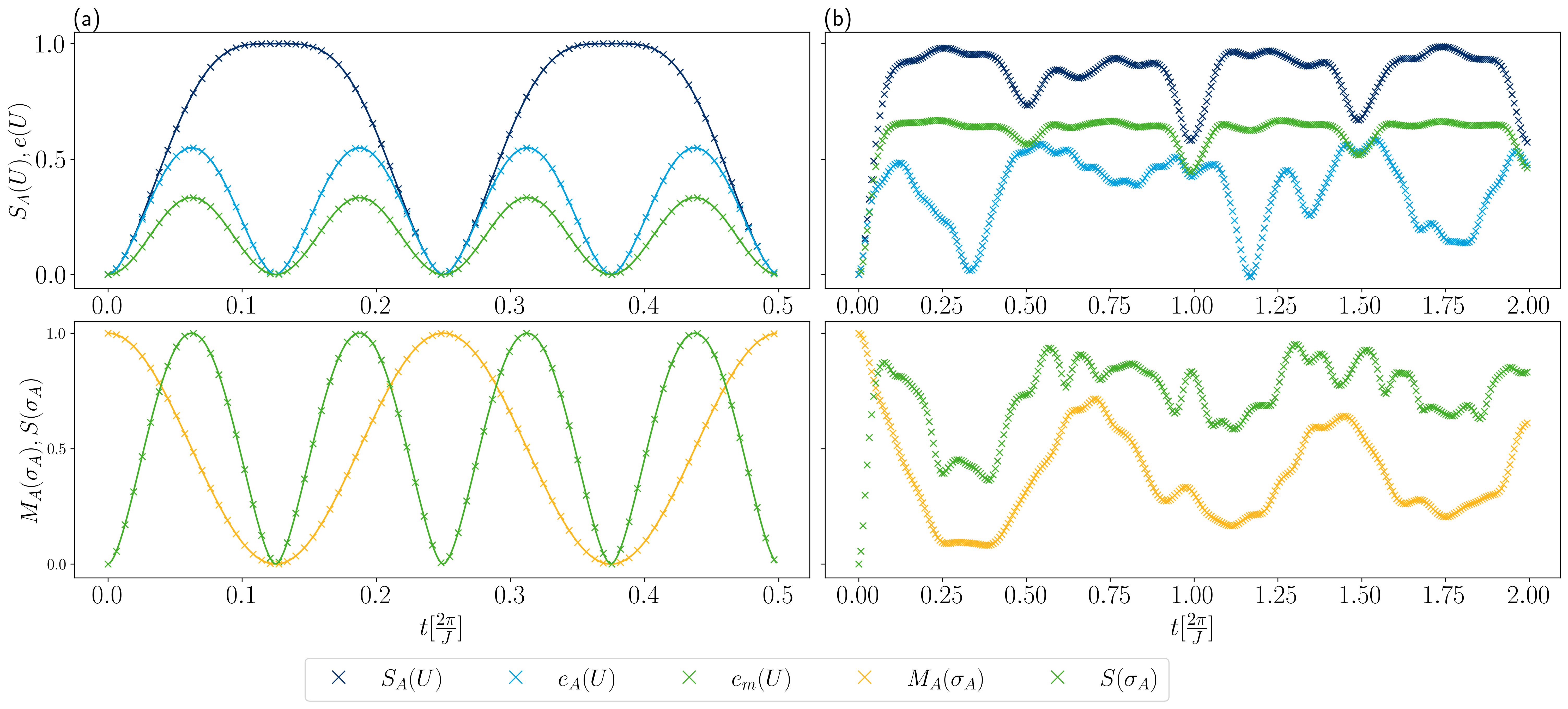}
	\caption{\textbf{Open quantum dynamics for the isotropic Heisenberg model of a subsystem of one of two neighboring qubits (left) and two qubits of a 2D grid of $3\times 2$ qubits (right).} The upper panel shows non-locality measure $S_A(U)$ and entangling power measures $e_A(U)$ and $e_m(U)$ (see App.~\ref{app:entanglement_gen}) for the respective time evolution operator, while the lower panel contains the total magnetization $M_A(\sigma_A)$ and the entanglement entropy $S(\sigma_A)$ of the time evolved state. The initial state of the total system reads $\ket{1}\otimes \ket{0}$ in (a) and $\ket{11}\otimes \ket{0000}$ in (b). In (a), solid curves represent analytical predictions derived in Eqs.~(\ref{eq:e1U} - \ref{eq:Sr1t}).}
	\label{fig:iso_Heisenberg_full}
\end{figure*}
Fig.~\ref{fig:iso_Heisenberg_full} repeats the numerical experiment shown in the main text, but also includes the entangling power measures $e_A(U)$ and $e_m(U)$ introduced in App.~\ref{app:entanglement_gen}. We compare the numerical results with the analytical derivation above and find an exact match. As we discuss parts of Fig.~\ref{fig:iso_Heisenberg_full} already in the main text. we focus on features of the entangling powers here. 

In the two-qubit case (cf. Fig.~\hyperref[fig:iso_Heisenberg_full]{\ref*{fig:iso_Heisenberg_full}(a)}), the non-locality undergoes four oscillation periods, representing the oscillation of $U$ between identity and swap operator. In accordance to this, the entangling powers both show an oscillation and vanish at the extreme points of the non-locality function $S_A(U)$. As they reach zero when $U$ equals the swap operator, while the non-locality stays maximal, they thus both undergo twice as many oscillations.

On a $3 \times 2$ qubit grid (cf. Fig.~\hyperref[fig:iso_Heisenberg_full]{\ref*{fig:iso_Heisenberg_full}(b)}), the entangling power measures no longer follow defined oscillations, but start off at zero, quickly rise and then stay at a non-zero value most of the time. $e_A(U)$, since it singles out swap operations, does become close to zero for certain simulation times. However, the trial state shows non-zero entanglement entropy at those times. As a consequence, $e_A(U)$ does not seem to be a good measure of entangling power when the dimensions $d_A$ and $d_B$ of subsystems $\mathcal{H}_A$ and $\mathcal{H}_B$ no longer match. The mean entanglement generation $e_m(U)$, on the other hand, is not in disagreement with this. Similar to the non-locality, it rises quickly, but stays around approximately $65 \%$ of its maximal value admitting dips at the same points as the non-locality $S_A(U)$, thus allowing for non-zero entanglement throughout the considered time interval.

\end{document}